\documentclass{aa}  
\usepackage{xspace}             
\usepackage{graphicx}
\usepackage{txfonts}
\usepackage{lipsum}
\usepackage{float}
\usepackage{subcaption}         
\usepackage{lscape}             
\usepackage{placeins}           

\usepackage[colorlinks=true,linkcolor=black,citecolor=blue,urlcolor=magenta]{hyperref}
\usepackage{siunitx}
\usepackage[dvipsnames, table]{xcolor} 

\usepackage[version=4]{mhchem}

\newcommand{\wis}{WISPIT\xspace} 
\newcommand{\wisa}{WISPIT~1\xspace}
\newcommand{\wisb}{WISPIT~1b\xspace}
\newcommand{\wisc}{WISPIT~1c\xspace}

\newcommand{\mura}{$\mu_{\alpha}$}
\newcommand{\mudec}{$\mu_{\delta}$}

\newcommand{\kms}{km\,s$^{-1}$}

\newcommand{\masyr}{\ensuremath{\rm mas\,yr^{-1}}}

\newcommand{\mj}{\ensuremath{\rm{M_J}}\xspace}

\newcommand{\msun}{\ensuremath{M_\sun}}

\newcommand{\EBV}{E\,($B-V$)}

\begin{document}

\title{WIde Separation Planets In Time (WISPIT):} 

\subtitle{Two directly imaged exoplanets around the Sun-like stellar binary WISPIT 1}

\titlerunning{Two directly imaged exoplanets around WISPIT 1}

\author{Richelle F. van Capelleveen\inst{1}
\and
Matthew A. Kenworthy\inst{1}
\and
Christian Ginski\inst{2}
\and
Eric E. Mamajek\inst{3}
\and
Alexander J. Bohn\inst{1}
\and
Rico Landman\inst{1,4}
\and
Tomas Stolker\inst{1}
\and
Yapeng Zhang\inst{5}
\and
Nienke van der Marel\inst{1}
\and
Ignas Snellen\inst{1}
}

\institute{Leiden Observatory, Leiden University, Postbus 9513, 2300 RA Leiden, The Netherlands\\
\email{capelleveen@strw.leidenuniv.nl}
 \and
School of Natural Sciences, Center for Astronomy, University of Galway, Galway, H91 CF50, Ireland
 \and
Jet Propulsion Laboratory, California Institute of Technology, 4800 Oak Grove Drive, M/S 321-162, Pasadena, CA, 91109, USA
 \and
SRON Netherlands Institute for Space Research, Niels Bohrweg 4, 2333 CA Leiden, The Netherlands
 \and
Department of Astronomy, California Institute of Technology, Pasadena, CA 91125, USA
 }

\date{Received July 24, 2025; accepted August 25, 2025}

 
  \abstract
   {Wide separation gas giant planets present a challenge to current planet formation theories, and the detection and characterisation of these systems enables us to constrain their formation pathways.}
   {The WIde Separation Planets In Time (WISPIT) survey aims to detect and characterise wide separation planetary-mass companions over a range of ages from $<5$ to $20$\,Myr around solar-type host stars at distances of 75-500 (median 140) parsecs.} 
   {The WISPIT survey carries out two 5 minute $H$-band exposures with the VLT/SPHERE instrument and IRDIS camera, separated by at least six months to identify co-moving companions via proper motion analysis.
   These two $H$-band observations in combination with a follow-up $K_s$-band observation were used to determine the colour-magnitude of the co-moving companions and to derive their masses by comparing to AMES-COND and AMES-DUSTY evolutionary tracks.}
   {We report the discovery of \wisb and \wisc, two gas giant exoplanets that are co-moving with the stellar binary \wisa, which itself consists of a K4 star and M5.5 star in a multi-decadal orbit.
   The planets are at projected separations of 338\,au and 840\,au and have masses of 10\,\mj and 4\,\mj, respectively.
   }
   {We identified two common proper motion planetary companions to a (previously unknown) stellar binary with a Sun-like primary.
   These targets are ideal for follow up characterisation with both ground and space-based telescopes.
   Monitoring of the orbit with the GRAVITY interferometer will place constraints on their eccentricity, and spectroscopic characterisation will identify the composition and metallicity, providing information on their formation pathways.}

\keywords{planets and satellites: detection --- planets and satellites: formation --- stars: individual: \wisa}
\maketitle

%

\section{Introduction}
\label{sec:introduction}
\begin{figure}
   \centering
   \includegraphics[width=\hsize]{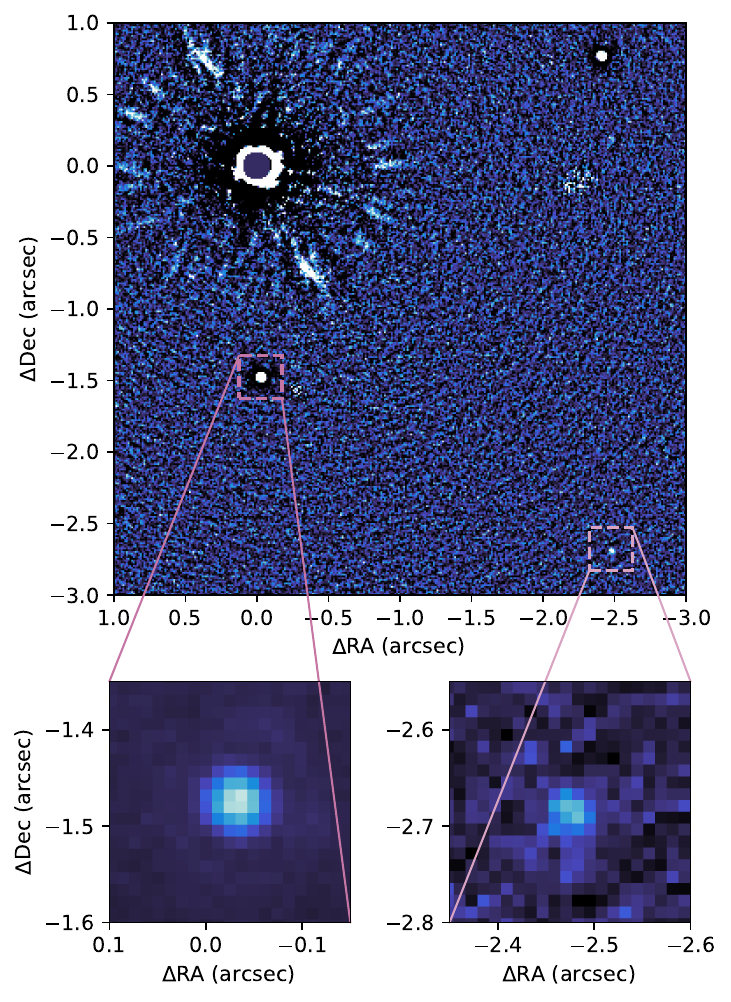}
      \caption{Detail of the \wisa system.
      The top panel shows the SPHERE/IRDIS $H$-band image taken on 2022 November 19, processed with unsharp masking to suppress the stellar halo. The stellar binary \wisa (behind the coronagraph) is located in the upper left, with its companions highlighted with coloured boxes. Zoomed-in images of the unmasked observation centred on companions \wisb and \wisc, are shown in the lower left and lower right panels, respectively.}
         \label{fig:wispit1_image}
\end{figure}
%
%
Gas giant planets are thought to form through either bottom-up or top-down formation mechanisms.
In bottom-up formation, an embryonic rocky protocore first forms and subsequently accretes gas to become a gas giant via core accretion \citep[CA; ][]{Pollack1996}.
While this is thought to be the dominant mode of planet formation, it becomes increasingly difficult for protocores to reach the critical mass required for gas accretion within the lifetime of the gas disk at separations much larger than the ice line \citep{Mordasini2012}.
Somewhat larger separations for \textit{in-situ} formation of gas giants can be reached through pebble accretion \citep[PA; ][]{Lambrechts2012}.
In contrast to classical CA, where the protocore grows by accreting planetesimals, with PA, the protocore grows via accretion of small solid particles that are drifting radially inward.
The efficiency of PA increases with stellar mass, and its relative efficiency compared to classical CA increases at larger separations, with the timescale to reach the critical core mass for gas accretion being shortened by a factor of 100 at 50~au \citep{Lambrechts2012}.
Nonetheless, the semi-major axis preference reported for gas giant formation via PA is on the lower end of the 10-100~au range \citep{Nielsen2019}.
Top-down formation mechanisms form a gas giant directly through gravitational collapse.
This can either happen by turbulent fragmentation processes of the collapsing proto-stellar cloud \citep{Kroupa2001} or by gravitational instability \citep[GI; ][]{Boss1997} where the gas giant forms from the direct collapse of part of the circumstellar disk \citep[see also review by ][]{Kratter16}.
Of top-down formation processes, the latter is thought to be the dominant planet formation mechanism, especially if the resulting planet is coplanar with the circumstellar disk and with other planets in the system.
Whereas gas giants represent the high-mass outcome of bottom-up formation, they represent the low-mass outcome of top-down formation \citep{Nielsen2019}, and where CA formation is more likely to occur at separations less than 35~au, GI formation is more likely to occur on wide orbits beyond 35~au \citep{Dodson-Robinson2009}.
%
%

To date, nearly a thousand of such gas giant planets with masses between 1 and 13 Jupiter masses have been detected and confirmed. 
Of these, fewer than 5\%\footnote{41 out of 992 as of July 22, 2025, retrieved from the NASA Exoplanet Archive \url{https://exoplanetarchive.ipac.caltech.edu}} have been discovered via direct imaging (DI), despite such planets being among the most promising targets for atmospheric characterisation \citep[e.g.,][]{Biller2018}.
This low detection rate is primarily due to challenges inherent to direct imaging: gas giants are intrinsically faint, and their signal is typically overwhelmed by the flux from the host star, especially at small angular separations.
Detecting them requires the use of starlight suppressing techniques and optics \citep[e.g.,][]{Kenworthy2025}, in combination with one or more post-processing techniques \citep[e.g.,][]{Claudi2024,Follette2023}.
Despite these challenges, several DI surveys over the past two decades have successfully revealed self-luminous gas giant planets at separations of several hundreds of au down to a few au from their parent stars: these include the Gemini Deep Planet Survey \citep{Lafreniere2007b}, GPIES \citep{Macintosh2014,Nielsen2019}, SPHERE SHINE \citep{Chauvin2017,Vigan2021}, SCExAO accelerating stars \citep{Currie2025,ElMorsy2024}, and the BEAST survey \citep{Delorme2024}.
Notable detections within 100 au of their host stars include the HR~8799 system \citep{Marois2008,Marois2010b}, $\beta~\mathrm{Pic\,b}$ \citep{Lagrange2009} and c \citep{Lagrange2019,Nowak2020}, 51~Eri~b \citep{Macintosh2015}, and more recently, AF~Lep~b \citep{Franson2023} and HD~135344~Ab \citep{Stolker2025}.
%
%

While direct imaging has revealed gas giant planets at a wide range of orbital separations, those detected at separations significantly larger than $100$~au are particularly intriguing as they present a challenge to planet formation theory.
Among these detections are HD~106906~b \citep{Bailey2014}, YSES~1b \citep{Bohn2019}  and c \citep{bohn2020b}, COCONUTS-2b \citep{Zhang2021} and b~Cen~b \citep{Janson2021}.
%
%
It is unclear whether such companions formed in situ through fragmentation or GI processes, or whether they formed closer to their host stars, either through top-down or bottom-up formation mechanisms, and were subsequently dynamically scattered to wide orbits.
Disentangling these scenarios requires comparing observations with planet formation models, simulations and predicted occurrence rates \citep{Mordasini2009a,Mordasini2009b,Forgan2018}.
Although tremendous effort and progress have been made, previous analyses have not yielded conclusive constraints on their formation pathways due to the small sample size of directly imaged wide separation gas giants \citep[e.g.,][]{Vigan2017,Bowler2018}.
One recent successful approach in expanding this sample size is the Young Suns Exoplanet Survey \citep[YSES; ][and van Capelleveen in prep.]{Bohn2021}, a VLT/SPHERE survey of a homogeneous sample of 70 young, solar-mass stars in the Lower Centaurus Crux (LCC) subgroup of the Scorpius-Centaurus association \citep[$\sim$15\,Myr, $\sim$120\,pc;][]{Pecaut2016}.
Among its findings is YSES 1 \citep{Bohn2020,bohn2020b}: the first multi-planet system around a young solar analogue.
%
%

The WIde Separation Planets In Time (WISPIT) survey extends the scope of YSES by targeting young Sun-like stars spanning a broader range of ages, and other regions of the sky.
To investigate whether wide orbit gas giants form in situ or form closer to their host star and are later scattered outward, it is necessary to sample stars from the earliest timescales of top-down formation (0.5~Myr) to the end of bottom-up formation (5-10~Myr).
This age-diverse sample of young, solar-type stars was constructed from the pre-main sequence catalogue of stars younger than 20~Myr compiled by \citet{Zari2018}.
This sample is magnitude-limited (G<13~mag) to enable AO-assisted imaging with VLT/SPHERE.
After applying selection criteria based on youth indicators, stellar mass, suitability for common proper motion analysis, and observability with VLT/SPHERE, we obtained a final sample of 178 solar-mass stars with ages ranging from $<5$ to 20 Myr (median: 8.5~Myr) and distances from 75 to 500\,pc (median: 140\,pc).
%
%

In this paper, we present the first discovery of this survey: \wisa. This young ($\sim$16 Myr) system located 229 parsecs away is a stellar binary with a solar-type primary, and hosts two comoving planetary mass companions at large semimajor axes, see Fig. \ref{fig:wispit1_image}.
After introducing our proposed nomenclature for the survey, we describe the observations (Sect.~\ref{sec:observations}) and the data reduction (Sect.~\ref{sec:data_reduction}). We present an analysis of the \wisa stellar components and the companions of the system in Sect.~\ref{sec:analysis}. In Sect.~\ref{sec:discussion}, we discuss the results and place them in the broader context of known systems. Finally, we summarise our findings and outline recommendations for future observations in Sect.~\ref{sec:conclusion}.

\section{Nomenclature of the \wis targets}
The \wis survey will have a dedicated catalogue that will be used for all targets in the survey. The WISPIT acronym has been accepted by the \textit{IAU Commission B2 Working Group on Designations} and has subsequently been registered in the dictionary\footnote{\url{https://cds.unistra.fr/cgi-bin/Dic-Simbad}} of the Simbad database \citep{wenger2000}. 
The nomenclature for the \wis acronym follows a 2x2 format, with two characters representing the star and two representing the planet. 
Here the host star is identified as WISPIT NNN where \textit{NNN} is an ordinal number with at most three digits. 
WISPIT NNNb is its planet, where \textit{b} is generic and may also be \textit{c, d,} etcetera. 
WISPIT NNNA is the host star that is itself one of a component of a common proper motion system (CPMS), where \textit{A} may also extend to \textit{B, C,} and so on. 
We note that in a CPMS with components \textit{A} and \textit{B}, by definition \textit{A} designates the brighter star of the system. 
If the stars in a CPMS are well separated or if the secondary component is significantly fainter than the primary, using \textit{NNNA} and \textit{NNNAb} is not necessary. 
Instead, \textit{NNN} and \textit{NNNb} can be used for the primary star and its planet respectively, with \textit{NNNB} assigned to a wide and/or faint stellar companion. 
As the survey contains 178 targets, the highest ordinal number assigned is 178; the final target in the survey will thus be named WISPIT 178. 
For instance, if this target were to be a stellar binary with a planet orbiting the secondary star, the primary would be named WISPIT 178A with secondary WISPIT 178B that has a planet WISPIT 178Bb.

Following the nomenclature described above, the first target in the survey is designated as WISPIT 1, with its planets WISPIT 1b and WISPIT 1c. Here WISPIT 1 refers to the primary star in the binary system, and WISPIT 1B denotes the significantly fainter secondary star.

\section{Observations}
\label{sec:observations}
\begin{table}
\caption{IDs, Astrometry and Photometry for \wisa}
\label{tab:star}
\def\arraystretch{1.2}
\setlength{\tabcolsep}{12pt}
\begin{tabular}{@{}lll@{}}
\hline\hline
Parameter & Value & Ref.\\ 
\hline
Gaia DR3 & 5517037434775143808 & (1)\\
2MASS & J07511168-5008158   & (2)\\
WISE & J075111.66-500815.8 & (3)\\
TIC & 268764100           & (4)\\
\hline
RA* $\alpha$  [deg] & 117.79855076351 & (1)\\
Dec* $\delta$ [deg] & -50.13768323600 & (1)\\
Parallax $\varpi$ [mas] & $4.3652\pm0.0137$ & (1)\\
Distance $d$ [pc] & $228.85^{+0.59}_{-0.68}$ & (5)\\
pmra $\mu_{\alpha}$ [mas/yr]  & $-19.824\pm0.016$ & (1)\\
pmdec $\mu_{\delta}$ [mas/yr] & $14.021\pm0.017$ & (1)\\
$v_r$ [km/s] & $21.08\pm5.61$ & (1)\\
$v_r$ [km/s] & $19.9\pm3.0$ & (6)\\
$v_r$ [km/s] & $19.0\pm3.0$ & (6)\\
\hline
$G$ [mag] & $12.722928\pm0.002907$ & (1)\\
$B_p - R_p$ & 1.427105 	& (1)\\
$B_p-G$ & 0.642953 & (1)\\
$G-R_p$ & 0.784152 & (1)\\
$B$ [mag] & $14.171\pm0.052$ & (7)\\
$V$ [mag] & $13.146\pm0.044$ & (7)\\
$B-V$ [mag] & $1.026\pm0.068$ & (7)\\
$g'$ [mag]  & $13.655\pm0.022$ & (7)\\
$r'$ [mag]  & $12.725\pm0.029$ & (7)\\
$i'$ [mag]  & $12.297\pm0.009$ & (7)\\
$J$ [mag]   & $10.859\pm0.026$ & (2)\\
$H$ [mag]   & $10.301\pm0.023$ & (2)\\
$K_s$ [mag] & $10.121\pm0.019$ & (2)\\
$W1$ [mag]  & $10.040\pm0.023$ & (8)\\
$W2$ [mag]  & $10.016\pm0.020$ & (8)\\
$W3$ [mag]  & $9.908\pm0.042$ & (8)\\
\hline
\end{tabular}
\tablefoot{*=ICRS, epoch J2016.0}
References:
(1) \citet{GaiaDR3},
(2) \citet{Cutri2003},
(3) \citet{Cutri2012},
(4) \citet{Stassun2019},
(5) \citet{BailerJones2021},
(6) \citet{Zerjal2021},
(7) \citet{APASSDR9},
(8) \citet{AllWISE}.
\end{table}
\begin{table*}
\caption{SPHERE/IRDIS observations of \wisa.}
\label{tab:obs_setup}
\centering
\def\arraystretch{1.2}
\setlength{\tabcolsep}{9pt}
\begin{tabular*}{\textwidth}{@{}llllllll@{}}
\hline\hline
Observation date & Filter & Coronagraph & NEXP$\times$NDIT$\times$DIT\ & $\omega$ & $ X$ & $\tau_0$ \\
(yyyy-mm-dd) & & & (1$\times$1$\times$s)  & (\arcsec) & & (ms)\\
\hline
2022-11-19 & $H$ &N\_ALC\_YJH\_S& 4$\times$2$\times$32  & $0.565\pm0.009$ & $1.116\pm0.001$ & $3.800\pm0.346$\\
2023-12-03 & $H$ &N\_ALC\_YJH\_S& 4$\times$2$\times$32  & $0.290\pm0.000$ & $1.110\pm0.000$ & $13.900\pm0.000$ \\
2024-11-30 & $K_s$ &N\_ALC\_Ks& 7$\times$1$\times$64  & $0.481\pm0.010$ & $1.151\pm0.003$ & $6.700\pm0.545$ \\
\hline
\end{tabular*}
\tablefoot{Observation setup and conditions for all \wisa observations. All filters are SPHERE broadband filters. The total integration time is the product of the number of exposures (NEXP), the number of subintegrations per exposure (NDIT) and the detector integration time (DIT). The seeing is denoted by $\omega$, the airmass by $X$ and the coherence time by $\tau_0$.}
\end{table*}
The combined stellar properties of WISPIT 1 are listed in Table~\ref{tab:star}.
\wisa has been observed on UTC~2022-11-19T08, UTC~2023-12-03T08, and UTC~2024-11-30T06 as part of the \wis survey in programs 110.23XJ.003, 112.25X3.003 and 114.27EK.003 respectively.
%
Obervations were taken with Spectro-Polarimetric High-contrast Exoplanet REsearch \citep[SPHERE;][]{Beuzit2019}, the coronagraphic facility mounted at the Nasmyth platform of 8.2m Unit Telescope 3 (UT3) of the Very Large Telescope (VLT).
To correct for atmospheric turbulence and internal defects, SPHERE utilises the extreme adaptive optics (AO) module called SPHERE AO for eXoplanet Observation \citep[SAXO;][]{Fusco2006,Fusco2014}.
All observations were conducted in Classical Imaging (CI) mode using the Infrared Dual-band Imager and Spectrograph \citep[IRDIS;][]{Dohlen2008} in pupil-stabilized mode.
The first two epochs use broadband filter $H$ and the third epoch uses broadband filter $K_s$.
To avoid saturation by the primary star in the science exposures, an apodized Lyot coronagraph was used to attenuate the stellar halo \citep{Soummer2005,Carbillet2011}.
An overview of the setup and observation conditions of the science observations is provided in Table \ref{tab:obs_setup}.
In addition to these science frames, we obtained center frames, sky frames and flux frames. The center frames are acquired by applying a sinusoidal pattern to the deformable mirror, creating a waffle pattern to determine the star's position behind the coronagraph.
The sky frames are taken at an offset position without AO correction and without any sources in the field of view, enabling the subtraction of instrumental and thermal background.
The flux frames serve as a photometric reference for detected point sources in the science frames, and are taken without a coronagraph.
In $H$-band, near-infrared neutral density filter ND1.0\footnote{For filter description and transmission, see \url{https://www.eso.org/sci/facilities/paranal/instruments/sphere/inst/filters.html}.} was used to allow for longer exposures without saturating the detector.

\section{Data Reduction} 
\label{sec:data_reduction}

All SPHERE observations have been (pre-)processed with (modified) PynPoint modules \citep{Amara2012,Stolker2019}.
This includes bad pixel removal, flatfielding, sky subtraction, anamorphic distortion correction, centering, derotating and median combining the exposures.
The anamorphic distortion \citep[see][]{Maire2016} is corrected by multiplying the y-axis by $1.0062\pm0.0002$, as stated in the SPHERE manual\footnote{SPHERE manuals: \url{https://www.eso.org/sci/facilities/paranal/instruments/sphere/doc.html}}.
The images are derotated to the parallactic angle on the sky and the static pupil offset of $135.99\pm0.11\deg$.
An additional rotation of $1.76\pm0.04\deg$ was applied to correct for the true north offset \citep{Maire2021}.
The pixel scales used for astrometric calibration are $12.246\pm0.009\,\si{mas.yr^{-1}}$ in $H$-band and $12.266\pm0.009\,\si{mas.yr^{-1}}$ in $K_s$-band based on the five-year analysis of SPHERE astrometric calibration data presented in \cite{Maire2021}.

Inspection of flux frames showed that an unknown observational issue caused all first flux frames of all three observations to be unusable due to lower signal and duplicated sources on the detector.
It appears to be specific to an attribute of the observation of this source, as it is not present in most observations of other WISPIT targets with the same observation setup.
To address this issue, in both $H$-band observations the first seven flux frames were discarded and in the $K_s$-band observation the first three flux frames were discarded.

The median-combined unsharp-masked image of the 2022 $H$-band exposures, highlighting the planetary-mass companions, is shown in Fig.~\ref{fig:wispit1_image}.
Annotated images of the median-combined observations from all epochs, along with a discussion of their noise properties, are included in Appendix~\ref{app:obs}.

\section{Results and Analysis}
\label{sec:analysis}

\wisa is a close stellar binary; we discuss its properties in Sect.~\ref{sec:binary_fit} and show that it exhibits negligible relative motion over our two-year baseline. 
The photometry and age of the primary are derived in Sect.~\ref{sec:age}.
In Sect.~\ref{subsec:astrometric_analysis}, we demonstrate that \wisb and \wisc share common proper motion with the star, and in Sect.~\ref{subsec:photometric_analysis} we show that their photometry is consistent with planetary mass objects.

\subsection{\wisa is a stellar multiple}
\label{sec:binary_fit}
\begin{table*}
\caption{Best-fit parameters of a synthetic binary model to \wisa}
\label{tab:binary_fit_parameters}
\centering
\def\arraystretch{1.2}
\setlength{\tabcolsep}{8.3pt}
\begin{tabular*}{\textwidth}{@{}llllllll@{}}
\hline\hline
 & & \multicolumn{3}{|c|}{primary} & \multicolumn{3}{c|}{secondary} \\
Observation date & Filter & Peak flux &  $\Delta x$ & $\Delta y$ & Peak flux &  $\Delta x$ & $\Delta y$ \\
(yyyy-mm-dd) &  & (ADU) & (pixels) & (pixels) & (ADU) & (pixels) & (pixels) \\
\hline
2022-11-19 & $H$    & $2998 \pm 23$ & $-0.357 \pm 0.204$ & $-0.418 \pm 0.204$ & $988 \pm 23$ & $1.536 \pm 0.204$ & $1.736 \pm 0.204$ \\
2023-12-03 & $H$    & $2908 \pm 28$ & $-0.391 \pm 0.204$ & $-0.585 \pm 0.204$ & $1170 \pm 28$ & $1.307 \pm 0.204$ & $1.972 \pm 0.204$ \\
2024-11-30 & $K_s$  & $7109 \pm 80$ & $-0.447 \pm 0.204$ & $-0.791 \pm 0.204$ & $3095 \pm 80$ & $1.387 \pm 0.204$ & $2.439 \pm 0.204$ \\
\hline
\end{tabular*}
\end{table*}

\begin{table*}
\caption{Colour and apparent magnitudes of WISPIT 1A and WISPIT 1B}
\label{tab:binary_colour_mag}
\centering
\def\arraystretch{1.2}
\setlength{\tabcolsep}{15pt}
\begin{tabular*}{\textwidth}{@{}llllll@{}}
\hline\hline
Star & $H$ & $K_s$ & $H-K_s$ & T & Spectral type \\
    & (mag) & (mag) & (mag) & (K) & \\
\hline
Primary (A) & $10.63 \pm 0.04$ & $10.51 \pm 0.02$ & $0.12 \pm 0.04$ & $4670^{+990}_{-540}$ & K4V (G5V–K7V) \\
Secondary (B) & $11.75 \pm 0.08$ & $11.42 \pm 0.03$ & $0.34 \pm 0.09$ & $2900^{+560}_{-280}$ & M5.5V (M2.5V–M7.5V) \\
\hline
\end{tabular*}
\end{table*}

\begin{figure*}[h!]
   \centering
   \includegraphics[width=0.75\textwidth]{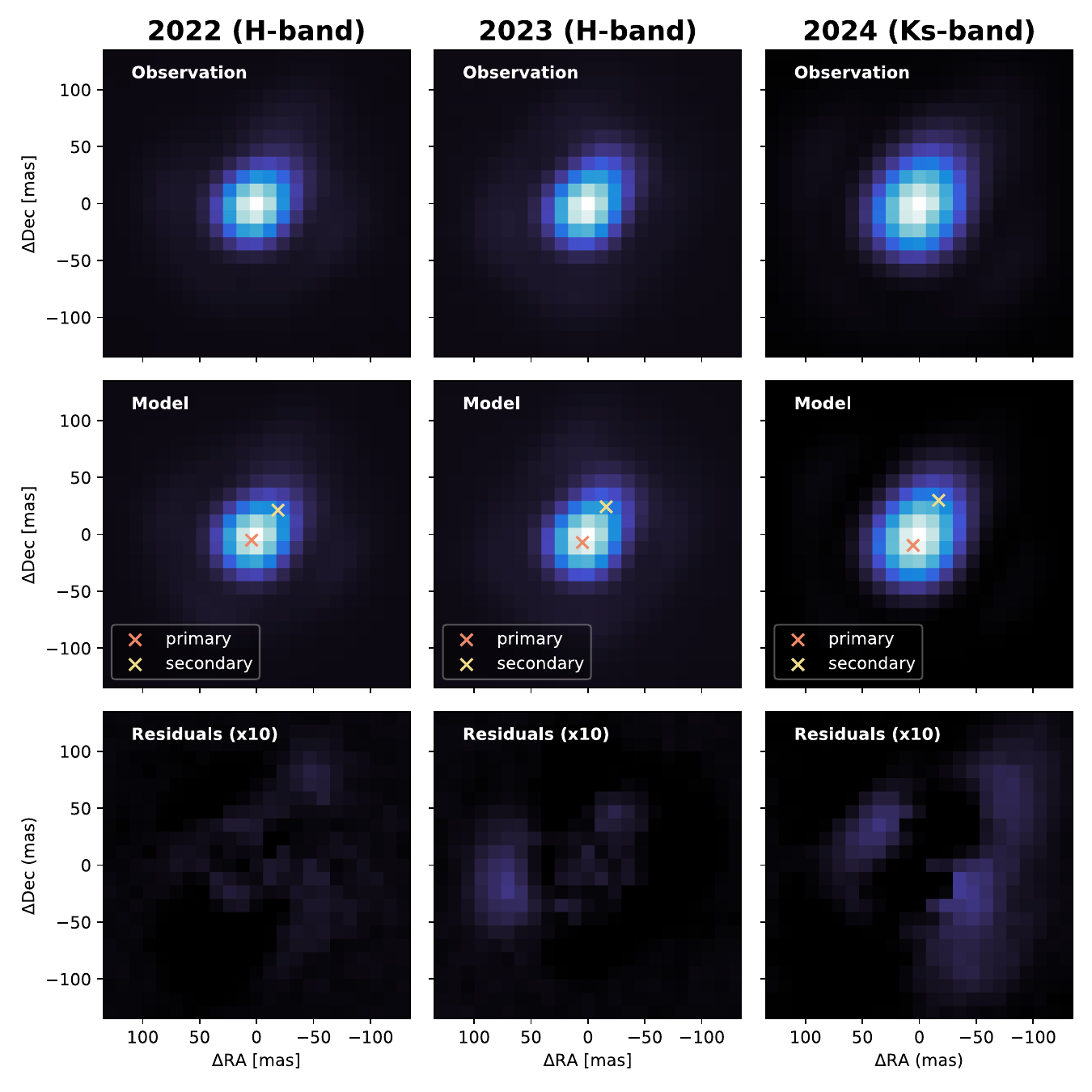}
      \caption{Synthetic binary model fit to the flux calibration images of \wisa over three different epochs. The top row shows the observation, the second row shows the best-fit synthetic binary model with best-fit positions of the primary star and secondary star marked with crosses, and the bottom row shows the residuals. The model and residuals inherit the colour scale from the corresponding observation, demonstrating the similarity in intensity between model and observation. The residual images are multiplied by a factor of 10, and show minor structures on the order of magnitude of less than $\sim$5\% of the peak flux of the observation.}
         \label{fig:binary_fit}
\end{figure*}

SPHERE flux calibration images of \wisa revealed it to be a previously unresolved binary (see Fig.~\ref{fig:binary_fit}). 
All three epochs show the presence of a marginally spatially resolved secondary companion to the north-west of the primary star.
To confirm that the secondary star is bound and shows minimal orbital motion over a period of two years, as well as to disentangle the flux of the primary star from that of the secondary star, we used a Markov chain Monte Carlo \citep[MCMC;][]{MacKay03} fitting routine to determine their relative positions and flux ratio. 
We constructed synthetic model binary stars from normalised median PSFs and used MCMC to obtain the best-fit model. 
To estimate the median PSF in $H$ and $K_s$ bands, we leveraged flux observations of single stars from YSES to construct normalised median PSFs, with peak flux scaled to unity. The sources used to create these PSFs are listed in Appendix \ref{app:median_psfs}. 
These median PSFs were then rotated according to the position angles of the flux frames of the \wisa observations.
To model the binary in the imaging data, we defined a log-probability function composed of a log-likelihood and log-prior term, evaluated using the \texttt{emcee} MCMC sampler \citep{ForemanMackey13}. 
The free parameters $\theta$ describe the offsets in $x$ and $y$ with respect to the centre of the frame and the peak fluxes of the primary and the secondary star.
The log-likelihood is computed as
\[
\ln \mathcal{L}(\theta) = -\frac{1}{2} \sum_{i} \left( \frac{D_i - M_i(\theta)}{\sigma_i} \right)^2,
\]
where $D_i$ and $M_i(\theta)$ are the observed and model pixel values respectively, $\sigma_i$ is the corresponding uncertainty, and the sum is performed over pixels selected in a mask around both stars. 
We assume Gaussian noise with a standard deviation of 5\% of the signal. 
The synthetic model image $M(\theta)$ is generated by placing the median PSFs at the positions and with peak fluxes specified by $\theta$. 
The log-prior constrains the stellar flux to be positive; it returns $\ln P(\theta) = 0$ when all fluxes are strictly positive, and $-\infty$ otherwise. The total log-probability is then given by
\[
\ln \mathcal{P}(\theta) = \ln P(\theta) + \ln \mathcal{L}(\theta),
\]
which is used to evaluate the posterior probability during sampling. 
We used $128$ walkers and $30000$ iterations, discarded the first 1500 steps and thinned every 250 steps. All free parameters converged to Gaussian distributions; the resulting parameters are listed in Table \ref{tab:binary_fit_parameters}. 
The fitting routine underestimates the positional errors, so instead we adopted errors in offset of 2.5 mas ($\sim$0.2~pixel), corresponding to the SPHERE centering precision. The error in the peak flux is derived from the root mean square (rms) of the residuals. 
Fig. \ref{fig:binary_fit} shows the observed image, the best-fit synthetic model image and the residuals for all three epochs.

The astrometric fits show negligible relative motion of the secondary to the primary star, inconsistent with the proper motion of a distant background source.
The interpretation of \wisa as a binary system is consistent with the Gaia DR3 Renormalised Unit Weight Error (RUWE) of 1.488, which also suggests that the source may be non-single. 
The measured positions of \wisa correspond to a binary with a physical projected separation of at least 10.5~au, which, assuming a circular Keplerian orbit, corresponds to a period of at least 34 years.
We conclude that this is a gravitationally bound secondary companion, and that its presence has a negligible effect on the observed motion of other sources in the field over a baseline of two years.
%

%
\subsection{Photometric analysis and age classification of the primary star}
\label{sec:age}
\begin{figure}
    \centering
    \includegraphics[width=\columnwidth]{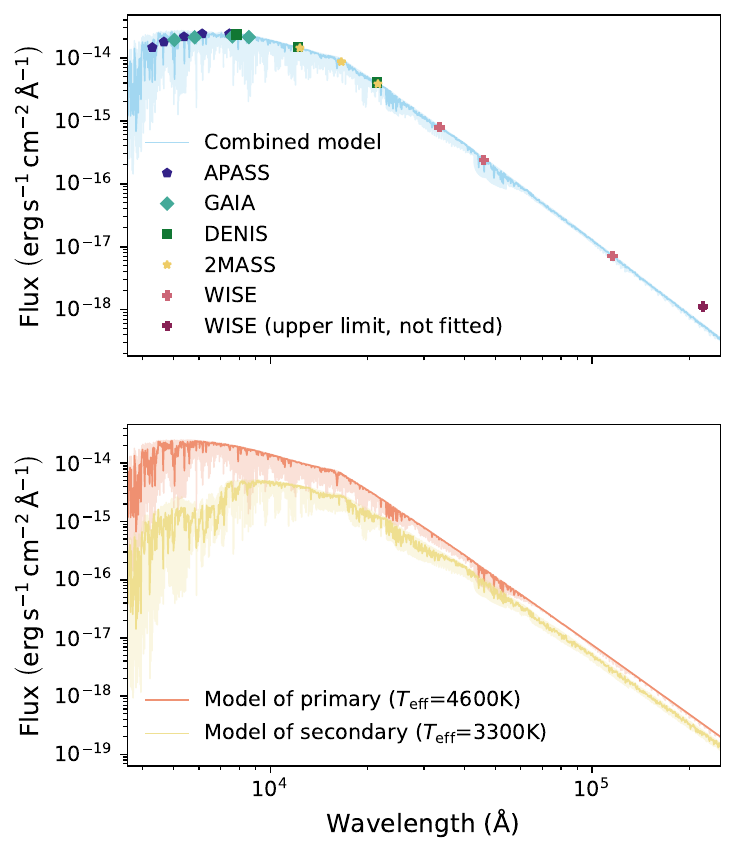}
    \caption{SED of \wisa with best-fit BT-Settle-CIFIST binary model ($\chi^2_r=5.1$). Spectra are presented as a low resolution (high opacity) version overlaid on the high resolution (low opacity) version. The top panel displays photometric data from various sources, represented by coloured markers, alongside the combined model spectrum of the primary and secondary components. The bottom panel presents the individual model spectra for the primary and secondary star. Best-fit parameters for the primary are $T_\mathrm{eff}=4600$\,K, $\log g=4.0$\,dex, and $L_\mathrm{bol}=0.36\,\mathrm{L_\odot}$. Best-fit parameters for the secondary are $T_\mathrm{eff}=3300$\,K, $\log g=4.5$\,dex, and $L_\mathrm{bol}=0.09\,\mathrm{L_\odot}$.}
    \label{fig:vosa_sed}
\end{figure}

The peak fluxes from the binary fit cannot be directly converted to stellar magnitudes. 
Moreover, we note that the $H$-band and $K$-band fluxes listed in Table \ref{tab:binary_fit_parameters} are not directly comparable due to attenuation in the $H$-band by an ND-filter. 
Instead, we used the flux ratio in combination with the 2MASS magnitudes (see Table \ref{tab:star}) of the unresolved system to derive individual magnitudes for the primary and secondary stars. 
Since the observed total flux in the $H$-band was higher in 2023 than in 2022, we used the weighted mean of the stellar fluxes across epochs to compute the $H$-band flux-ratio. 
The derivation of the individual apparent magnitudes is included in Appendix \ref{app:mag_decomposition}.
The resulting $H-K_s$ colour was used to estimate the effective temperature and spectral type based on Table 5 of \cite{Pecaut2013}\footnote{Spectral table from \url{https://www.pas.rochester.edu/~emamajek/EEM_dwarf_UBVIJHK_colors_Teff.txt} version 2022-04-16.}. 
The derived apparent magnitudes, colour, effective temperature and estimated spectral type are presented in Table \ref{tab:binary_colour_mag}. 

We performed a $\chi^2$ fit of a total of 19 photometric points from 2MASS \citep{Cutri2003}, WISE \citep{Cutri2012}, DENIS \citep{Epchtein1999}, NEOWISE \citep{Mainzer2014}, APASS DR9 \citep{APASSDR9}, and Gaia DR3 \citep{GaiaDR3} to synthetic models from BT-Settl CIFIST \citep{Allard2013} with Virtual Observatory SED Analyzer \citep[VOSA;][]{Bayo2008}.
We constrained the extinction to $A_v=0.047$ as derived in Appendix \ref{app:reddening_extinction}.
The temperatures were constrained based on the derived ranges presented in Table~\ref{tab:binary_colour_mag}.
The fit parameters of the primary were constrained to be in ranges $4300\leq T_\mathrm{eff} \leq 5300$ and $4\leq\log g\leq5$, and the fit parameters of the secondary were constrained to be in ranges $2600\leq T_\mathrm{eff} \leq 3400$ and $4\leq\log g\leq5.5$.
The best-fit consistent with the observed flux ratios in $H$- and $K_s$-band is presented in Fig.~\ref{fig:vosa_sed}.

\cite{Zerjal2021} report two equivalent width measurements of lithium absorption lines of \wisa measured with the Wide Field Spectrograph \citep[WiFeS;][]{Dopita2007}.
We computed the weighted mean using the inverse variance derived from the signal-to-noise ratio of each measurement, resulting in an equivalent width of EW(Li)=$0.325\pm0.009\,\mathrm{\AA}$.
We adapted the isochrones from \citet{Zerjal2021} to $H-K_s$ colour using Table 5 from \citet{Pecaut2013}; these are presented in Fig. \ref{fig:isochrones} in Appendix \ref{app:isochrones}.
As supported by the SED fit presented in Fig.~\ref{fig:vosa_sed}, the flux at the \ion{Li}{I}~6708~$\AA$ line is dominated by the primary star; the secondary star is not expected to contribute to the lithium absorption measurement.
Therefore, in our calculations of the age of \wisa, we assume that the equivalent width measurement is solely due to the primary star. 
The age is sampled by interpolating samples from the $H$, $K_s$, and EW(Li) distributions to the model isochrone grid presented in Fig.~\ref{fig:isochrones}.
The resulting sampled age is $15.6_{-1.2}^{+1.4}$~Myr, consistent within uncertainties with the reported age of $16.8^{+3.5}_{-3.9}$~Myr \citep{Kerr2021} for comoving co-distant star 2MASS J07485619-
4656229 (see Appendix \ref{app:environ}).
However, these errors are inferred only from the uncertainty on the EW(Li) measurement itself and the uncertainty on the photometry.
In reality, the cosmic scatter in lithium abundances would dominate the error on the age, so the formal errors on the age are likely underestimated.
To provide a more conservative estimate, and to bring the uncertainties in line with those typically reported in literature for age determinations (including that of the comoving companion), we inflate the age uncertainties by a factor of three and adopt an age of $15.6_{-3.7}^{+4.1}$~Myr.
%
%

\subsection{Astrometric analysis}
\label{subsec:astrometric_analysis}

\begin{figure*}
    \centering
    \begin{subfigure}{0.481\textwidth}
        \centering
        \includegraphics[width=0.99\textwidth]{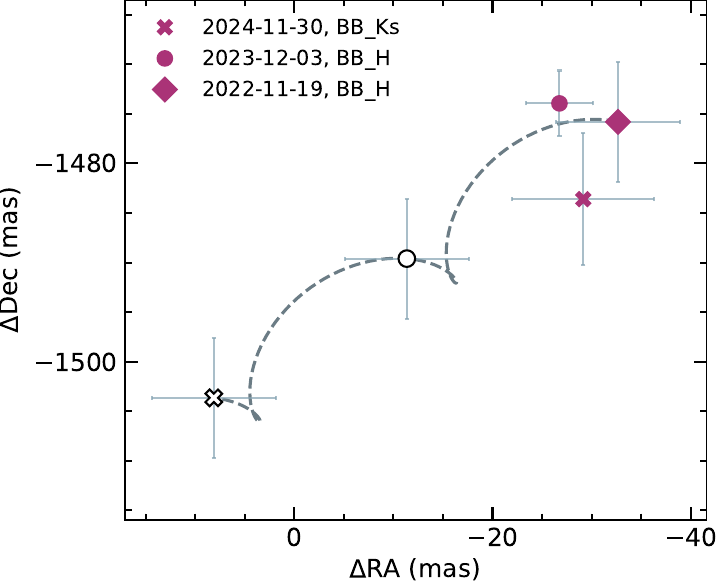}
        \caption{\wisb}
    \end{subfigure}%
    ~ 
    \begin{subfigure}{0.499\textwidth}
        \centering
        \includegraphics[width=0.99\textwidth]{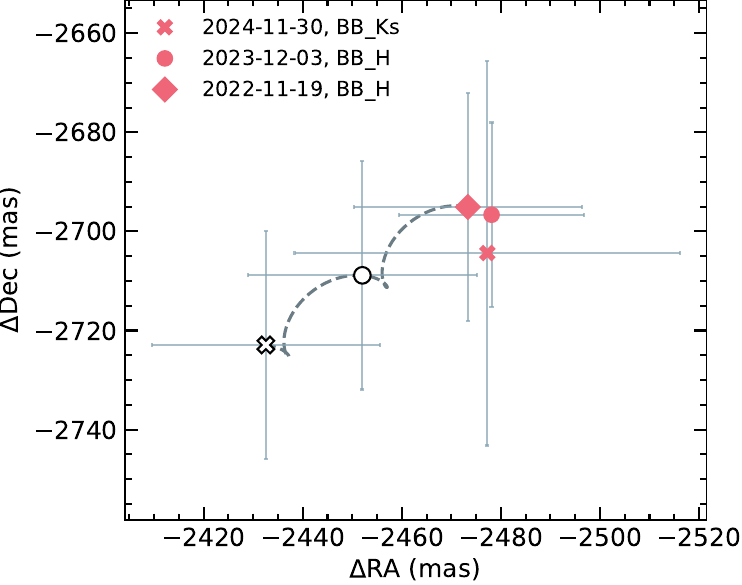}
        \caption{\wisc}
    \end{subfigure}
    \caption{Proper motion analysis of \wisb and \wisc. Each epoch is marked with a unique marker shape; diamond (2022), circle (2023) and cross (2024). The coloured version of the marker indicates the measured position of the companion. The unfilled (black outline, white center) version of this marker indicates the position that the companion would have been at at the corresponding date if it were a background object. The dashed curve depicts the parallactic track of a background object from first epoch to last epoch. The $\chi^2$ between the measured positions and the background track is 78.2 for \wisb and 3.8 for \wisc.}
    \label{fig:pm_plots}
\end{figure*}

\begin{table*}
\caption{Astrometry of \wisb and \wisc.}
\label{tab:astrometry_both}
\centering
\def\arraystretch{1.2}
\setlength{\tabcolsep}{18.97pt}
\begin{tabular*}{\textwidth}{@{}lllllll@{}}
\hline\hline
 & & \multicolumn{2}{|c|}{1b} & \multicolumn{2}{c|}{1c} \\
Observation date & Filter & Separation &  Position angle & Separation &  Position angle \\
(yyyy-mm-dd) &  & (\arcsec) & (\si{\degree}) & (\arcsec) & (\si{\degree}) \\
\hline
2022-11-19 & $H$ & $1.476\pm0.006$ & $181.27\pm0.24$ & $3.658\pm0.023$ & $222.54\pm0.36$\\
2023-12-03 & $H$ & $1.474\pm0.003$ & $181.04\pm0.13$ &$3.662\pm0.019$ & $222.58\pm0.29$\\
2024-11-30 & $K_s$ & $1.484\pm0.007$ & $181.13\pm0.28$ & $3.667\pm0.038$ & $222.49\pm0.61$\\
\hline
\end{tabular*}
\end{table*}

\wisb and \wisc are confirmed bound companions to \wisa based on common proper motion. Fig. \ref{fig:wispit1_image} indicates the positions of both planets, which shows that they both lie beyond the extent of the stellar point spread function (PSF) halo. 
This enabled astrometric measurements to be obtained by fitting a 2D Gaussian to the approximate positions of the companions in the median combined science frames. 
The fitting method used to fit the \texttt{Astropy} \citep{Astropy2013,Astropy2018,Astropy2022} 2D Gaussian model to the estimated positions was the \texttt{TRFLSQFitter}, a Trust Region Reflective algorithm with bound constraints and least-squares statistics. 
The fitting was constrained such that the centre of the Gaussian must lie within 5 pixels in both $x$ and $y$ direction of the manually estimated companion position. 
As an additional constraint, the Gaussian standard deviation was limited to be within 2.0 pixels of the median PSF standard deviation in the corresponding photometric band to ensure that the fitted full width at half maximum (FWHM) is consistent with that of SPHERE observations. 
The median PSF used to determine these constraints is constructed from observations from the Young Suns Exoplanet Survey \citep[YSES;][ and van Capelleveen in prep.]{Bohn2020} and is described in detail in Appendix \ref{app:median_psfs}.

The uncertainties in pixel positions are taken from the covariance matrix of the Gaussian fit. In conversion of pixel positions to separation (arcseconds) and position angle (degrees), uncertainties in pixel scale, true north correction, and pupil offset---as detailed in Sect.~\ref{sec:data_reduction}---are taken into account. 
Additionally, the centering precision of 2.5 mas of positioning the star behind the coronagraph is included in the error budget.
%
The resulting positions for \wisb and \wisc are listed in Table \ref{tab:astrometry_both}. 
Using the Gaia DR3 distance to \wisa (see Table~\ref{tab:star}) this places \wisb and \wisc at projected physical separations of 338 au and 840 au respectively. 
We present the measured positions of the companions alongside predicted background tracks for stationary background objects in Fig. \ref{fig:pm_plots}, and used \texttt{backtracks}\footnote{\url{https://github.com/wbalmer/backtracks}} \citep{backtracks_zenodo} to calculate the $\chi^2$ between the measured positions and the stationary background tracks.
\wisb has a $\chi^2$ of 78.2, reflecting the visual disagreement with the stationary background track.
\wisc has a relatively small $\chi^2$ of 3.8.
This low value is primarily driven by the large astrometric uncertainties, as the clustering of the astrometric measurements around the first epoch position is visually similar to that of \wisb. 
For comparison, the background sources in the field of view (see Fig. \ref{fig:background_ppm} in Appendix~\ref{app:background_ccs}) show better agreement with the stationary background tracks, yielding $\chi^2$ values of 10.6, 1.1 and 32.2. 
The latter higher value is attributable to a discrepancy in the position of the third epoch, which we suspect may be caused by binarity.
Unlike these background sources, which visually follow the predicted background tracks, both 1b and 1c show minimal motion with respect to the star.
Given their large projected physical separations from the host, this negligible observed motion over a two-year baseline is consistent with their interpretation as wide-orbit bound companions.



\subsection{Photometric analysis}
\label{subsec:photometric_analysis}

\begin{table*}
\caption{Photometric statistics and flux contrast of \wisb and \wisc.}
\label{tab:photometry_stats_ap_rad}
\centering
\def\arraystretch{1.2}
\setlength{\tabcolsep}{8.27pt}
\begin{tabular*}{\textwidth}{@{}lllllllll@{}}
\hline\hline
 & & & \multicolumn{3}{|c|}{1b} & \multicolumn{3}{|c|}{1c} \\
Observation date & Filter & Ap. rad. & SNR & FPF & Flux contrast & SNR & FPF & Flux contrast \\
(yyyy-mm-dd) & & (pix) & & & & & & \\
\hline
2022-11-19 & $H$ & 8.47 & 27.0 & $9.9 \times 10^{-27}$ & $(3.8 \pm 0.7) \times 10^{-4}$ & 5.6 & $1.3 \times 10^{-7}$ & $(2.3 \pm 0.5) \times 10^{-5}$ \\
2023-12-03 & $H$ & 8.47 & 66.8 & $2.3 \times 10^{-35}$ & $(3.7 \pm 0.7) \times 10^{-4}$ & 4.8 & $3.3 \times 10^{-6}$ & $(1.3 \pm 0.4) \times 10^{-5}$ \\
2024-11-30 & $K_s$ & 10.27 & 53.8 & $2.7 \times 10^{-24}$ & $(8.6 \pm 0.4) \times 10^{-4}$ & 1.7 & $4.3 \times 10^{-2}$ & $(7.1 \pm 1.2) \times 10^{-5}$ \\
\hline
\end{tabular*}
\tablefoot{Photometric detection statistics for both companions. The aperture radius (Ap. rad.) is the radius used to extract the photometry of the sources. The confidence of the detection is denoted by the signal-to-noise ratio as described by \citet{Mawet2014} and FPF denotes the false positive fraction. The flux contrast with \wisa is corrected for neutral density filters where relevant.}
\end{table*}

\begin{table*}
\caption{Apparent and absolute magnitudes, and colours of \wisb\ and \wisc}
\label{tab:wisb_wisc_colour_mag}
\centering
\def\arraystretch{1.2}
\setlength{\tabcolsep}{22pt}
\begin{tabular*}{\textwidth}{@{}lllllll@{}}
\hline\hline
Component & $m_{H}$ & $M_{H}$ & $m_{K_s}$ & $M_{K_s}$ & $H-K_s$ \\
         & (mag) & (mag) & (mag) & (mag) & (mag) \\
\hline
\wisb & $18.87^{+0.35}_{-0.27}$ & $12.07^{+0.35}_{-0.27}$ & $17.78^{+0.05}_{-0.05}$ & $10.98^{+0.05}_{-0.05}$ & $1.09^{+0.31}_{-0.23}$ \\
\wisc & $22.16^{+0.46}_{-0.32}$ & $15.37^{+0.46}_{-0.32}$ & $20.49^{+0.20}_{-0.17}$ & $13.69^{+0.20}_{-0.17}$ & $1.69^{+0.31}_{-0.24}$ \\
\hline
\end{tabular*}
\end{table*}

 \begin{figure}
    \centering
    \includegraphics[width=\hsize]{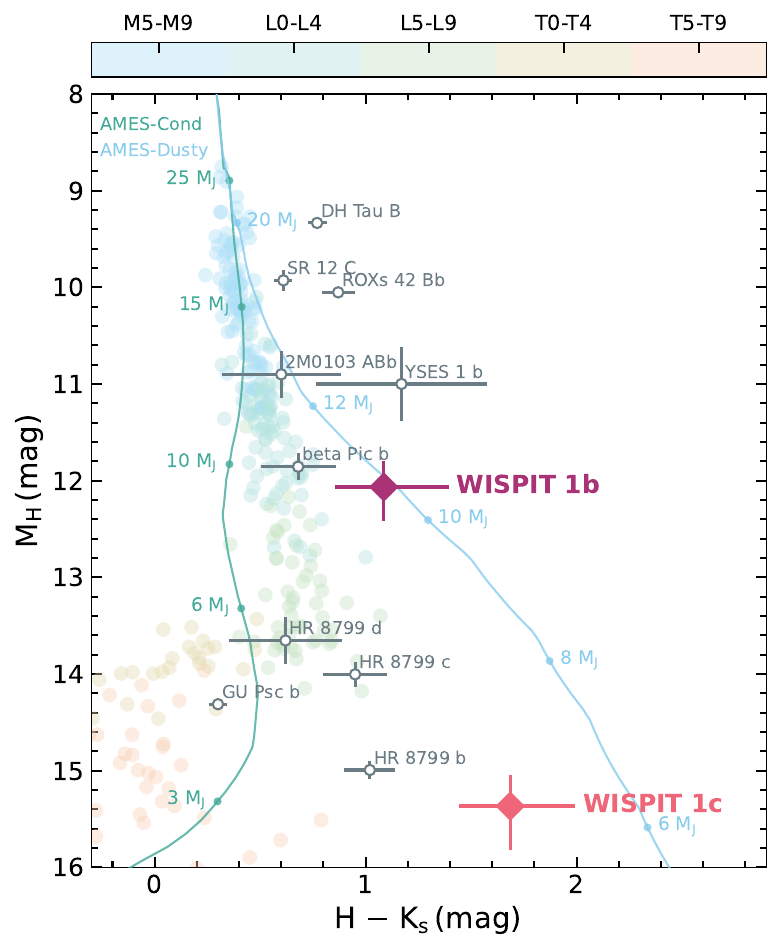}
    \caption{Colour-magnitude diagram of \wisb and \wisc, with field brown dwarfs of various spectral types and confirmed planetary companions. Teal and cyan tracks show 15.6\,Myr AMES-COND and AMES-DUSTY isochrones respectively. \wisb/\wisc are marked in purple/pink.}
    \label{fig:cmd}
    \end{figure}
To assess whether the photometry of the bound companions is consistent with that of planetary-mass objects, we extracted photometry in two broadband filters, $H$ and $K_s$, and fitted isochrones to estimate their mass. 
Considering the large angular separation between the companions and the star, both companions are in the background-limited regime of the observations, which allowed for using background-subtracted aperture photometry for flux extraction.
%

To derive the magnitudes of the companion candidates, we utilised the 2MASS magnitudes of the host star listed in Table \ref{tab:star}. 
The binary is not resolved by 2MASS; their reported magnitudes correspond to the combined flux of the primary and the secondary star. 
To ensure consistency when computing the magnitude contrast, we used an aperture that is large enough to encompass the majority of the flux from both the primary and the secondary star. 
Specifically, we adopted an aperture with a radius of twice the FWHM of the median PSF in that photometric band (see Appendix~\ref{app:median_psfs}), resulting in radii listed in Table \ref{tab:photometry_stats_ap_rad}. 
The flux frames were already centred on the binary by the PynPoint reduction routine, but more precise centering for aperture photometry was done by fitting a 2D Gaussian with the same fitting routine as described in Sect.~\ref{subsec:astrometric_analysis}.
The centre was constrained to lie within 5 pixels of the centre of the frame and the standard deviations of the fit were bounded to $\pm1.5$ times the standard deviation of the median PSF. 
These constraints were not actively limiting in any frame, and served primarily as a safeguard.
The final stellar flux was computed as the median of the flux from the individual frames, with uncertainties given by the standard deviation of the flux from these frames scaled by the ratio of the FWHM standard deviation to the FWHM mean across the frames.
In both photometric bands, the stellar flux was scaled to the science frames by applying a scale factor that accounts for the exposure time difference.
In $H$-band, the measured stellar flux was also attenuated due to use of an ND filter. 
To account for this we scaled the extracted flux to the science frames by using a factor that combined the exposure time ratio and the ND filter transmission.
Here the ND filter transmission was calculated as the weighted average of the ND filter's transmission across the broadband filter profile. 
The total scaled stellar flux and uncertainties are presented in Table \ref{tab:binary_fit_parameters}.

The flux from the companions in the field was extracted by using an aperture of the same radius as used for the stellar aperture (see Table \ref{tab:photometry_stats_ap_rad}), centred on the companion position determined in astrometric analysis (Sect.~\ref{subsec:astrometric_analysis}). 
For noise estimation, we used a modified version of the PynPoint module \texttt{falsealarm}, which estimates background flux by selecting background apertures of equal radius located at the same separation from the star. 
The two apertures closest to the companions were excluded to avoid contamination. 
To mitigate the influence of bad pixels and/or field sources in the noise apertures, as an adjustment to the module we applied \texttt{astropy.stats} sigma-clipping with a $2\sigma$ threshold and five iterations to the background apertures.
The noise statistics are given by the standard deviation of the integrated flux in the noise apertures multiplied with a correction factor to account for small sample statistics, following the methods from \citet[Eq.~8]{Mawet2014}.
For each companion, the signal, signal-to-noise ratio, false positive fraction and flux contrast with the binary were computed as weighted means from individual frame measurements. These results are summarised in Table~\ref{tab:photometry_stats_ap_rad} and were used in further analysis.

The flux contrast with the binary was converted to magnitude contrast. The apparent magnitude was then derived by assuming that the total measured stellar flux corresponds to the 2MASS catalogue value for the host star.
Absolute magnitudes for the companions were calculated by using the Gaia DR3 derived distance to the system (Table \ref{tab:star}). 
%
Apparent and absolute magnitudes of the companions are listed in Table \ref{tab:wisb_wisc_colour_mag}.
We note a slight discrepancy in the $H$-band magnitude between the 2022 and 2023 epochs for both companions, as well as all other field sources, with a lower magnitude (i.e., brighter) measured in 2023. 
This can be attributed to a higher measured stellar flux in 2023, resulting in a lower flux contrast of all candidate companions.
Additionally, we note that the signal and signal-to-noise are lower in $K_s$-band than in $H$-band, especially for \wisc. 
This is due to a higher background noise overall, as well as a thermal background pattern typical of SPHERE $K_s$-band observations, which is especially prominent near the position of 1c on the detector (see Fig.~\ref{fig:full_2024} in Appendix~\ref{app:obs}).
The colour-magnitude of the companions, along with 15.6~Myr AMES-COND and AMES-DUSTY isochrones \citep{Allard2001,Chabrier2000} and other confirmed planets with available magnitudes in both $H$ and $K_s$-bands, is shown in Fig. \ref{fig:cmd}. 
\wisc is among the faintest exoplanets discovered with in these bands, which is especially astonishing considering the confirmation required only two short exposures.
However, we note its similarity to YSES 1c \citep{bohn2020b}, of which no $K_s$ measurement exists.
The masses of the companions are estimated by interpolating to AMES models retrieved in colour-magnitude space with \texttt{species} \citep{Stolker2020b}.
We sampled from the companions' respective asymmetric $H$-magnitude, $K_s$-magnitude and age distributions, and interpolated the $H$-magnitude to the model isochrone grid. 
Here the age is the derived age of $15.6_{-3.7}^{+4.1}$~Myr.
This was done for both AMES-COND and AMES-DUSTY evolutionary models---a final mass was obtained by interpolating the $H-K_s$ colour between the two. 
The resulting masses are shown in Table \ref{tab:wisb_wisc_mass_models}.
The masses derived from both models, as well as the final adopted masses of $10.4^{+0.7}_{-0.8}$ $\mathrm{M_J}$ and 
$5.3^{+0.8}_{-0.6}$ $\mathrm{M_J}$ for \wisb and \wisc respectively, are consistent with the planetary-mass regime.

\begin{table}
\caption{Masses of \wisb and \wisc.}
\label{tab:wisb_wisc_mass_models}
\centering
\def\arraystretch{1.2}
\setlength{\tabcolsep}{29pt}
\begin{tabular}{@{}lll@{}}
\hline\hline
Model & Mass 1b & Mass 1c \\
    & ($\mathrm{M_J}$) & ($\mathrm{M_J}$) \\
\hline
AMES-COND & $9.24^{+0.86}_{-0.91}$ & $2.91^{+0.32}_{-0.34}$\\
AMES-DUSTY & $10.53^{+0.65}_{-0.76}$ & $6.24^{+0.67}_{-0.54}$\\
Interpolated & $10.4^{+0.7}_{-0.8}$ & $5.3^{+0.8}_{-0.6}$\\
\hline
\end{tabular}
\tablefoot{Mass estimates for both companions using AMES-COND and AMES-DUSTY evolutionary tracks. The final mass displayed in the Interpolated row is obtained by interpolating the H-Ks colour between the AMES-COND and AMES-DUSTY isochrones.}
\end{table}


\section{Discussion}
\label{sec:discussion}
\subsection{The stellar binary \wisa}
Our flux calibration observations revealed that \wisa is a close stellar binary. 
However, it shows negligible relative motion over the two-year baseline of our observations, indicating that the binarity does not significantly impact the relative astrometry of other sources in the field.
This allows us to treat the system effectively as a single astrometric reference for the purpose of tracking candidate companions.

We characterised the components of the \wisa system by deriving flux ratios in the $H$ and $K_s$-bands, and estimated the age of the primary by comparing its $H-K_s$ colour and lithium equivalent width to the model isochrones from \citet{Zerjal2021}.
While we addressed the considerable cosmic scatter in lithium abundances by adopting more conservative uncertainties on the age, we acknowledge that other, unquantified sources of uncertainty remain. 
Uncertainties stemming from disentangling the primary's flux from that of the secondary may be underestimated, and affect both the colour and the inferred age of the star.
Nevertheless, given the consistency with the age of the comoving star, we consider the age estimate to be reasonable.
High-resolution spectroscopic follow-up is required to better characterise the individual components of the binary system, and to provide a more robust constraint on its age.

\subsection{Planetary companions to \wisa}
Both \wisb and \wisc exhibit common proper motion with \wisa, which supports their interpretation as bound planetary companions.
The astrometric uncertainties on \wisc, as shown in Fig.~\ref{fig:pm_plots}b, are relatively large, and the possibility that it is a background object cannot be ruled out based on the current data, but the clustering of the astrometric measurements suggest that the positional uncertainties on \wisc may be overestimated, potentially resulting from its lower SNR and the higher background RMS at this large radial separation.
%
%
Follow-up observations are required to confirm its status as a bound companion.

This caution is further warranted by the recent case of YSES~2b, which was initially identified as a planetary companion to YSES~2, but with multiple SPHERE and GRAVITY epochs was later revealed to be a late type M dwarf star at a distance of over 2~kpc (Kenworthy et al. 2025, submitted). 
%
%
Such a scenario is a statistically unlikely event, and in the case of \wisa, this improbability is further reinforced by the fact that all other background sources in the field of view are consistent with stationary background sources.
It would therefore be improbable that both \wisb and \wisc are unrelated background sources that coincidentally move at the same apparent proper motion and in the same direction as \wisa.
Nevertheless, considering wide-separation planets exhibit little to no orbital motion over the baseline of a few years, this scenario is technically possible for all directly imaged wide-separation planets.
This highlights the importance of long-term astrometric monitoring with high precision instruments such as GRAVITY.

Despite the aforementioned caveats, the current data provides considerable robust evidence that both \wisb and 1c are indeed planetary-mass companions.
%
While the $K_s$-band detection of \wisc has a significance just under 5$\sigma$, making the $H-K_s$ colour and therefore the final interpolated mass somewhat uncertain, its mass derived from the $H$-band magnitude alone is in any case between $\sim$2.9~\mj (AMES-COND) and $\sim$6.2~\mj (AMES-DUSTY), placing it well within the planetary-mass regime.
With a mass of $10.4^{+0.7}_{-0.8}$~\mj, \wisb is well below the deuterium-burning limit, and its colour is comparable to that of other directly-imaged gas giants.

The large angular separation from \wisa makes both companions excellent targets for future photometric and spectroscopic follow-up observations, as this large distance limits contamination from the diffraction halo of the host star.
For \wisb, atmospheric characterisations can be performed with ground-based telescopes, while \wisc, due to its faintness, would require space-based instruments such as JWST to resolve spectroscopic features.
Astrometric and spectroscopic follow-up of both \wisb and 1c will be essential for conclusively validating their planetary nature and for constraining their compositions.

\subsection{Comparison to other systems}
\label{subsec:discussion_comparison}
The newly discovered \wisa system contributes to a growing population of wide-separation planetary systems, and shows similarities to existing systems, both in terms of system architecture and in terms of individual companion properties.
As can be seen in the CMD presented in Fig.~\ref{fig:cmd}, \wisb's photometry is comparable to that of $\beta~\mathrm{Pic\,b}$ \citep{Lagrange2009}. 
Though \wisa may be slightly younger, both systems are roughly similar in age and the mass of \wisb is comparable to that of $\beta~\mathrm{Pic\,b}$ \citep[$11\pm2$\,\mj,][]{Snellen2018}.
However, $\beta~\mathrm{Pic}$ is a more massive A-type star \citep[e.g.,][]{Gray2006}, and its companions orbit at much smaller separations.
\wisb also shows striking similarities to 1RXS~J160929.1-210524\,b, a  similar mass (7-12\mj) companion orbiting at a similar distance ($\sim$330~au) from its host star \citep{Lafreniere2008,Ireland2011,Lachapelle2015}. 
The star, located in the Upper Scorpius OB association, is roughly solar mass, but with an estimated age of $\sim$5~Myr is somewhat younger than \wisa.
Another $\sim$5~Myr star in that same region, GSC~06214-00210, hosts a more massive companion of $14-17$~\mj at the same projected separation of $\sim$330~au \citep{Kraus2008, Ireland2011,Lachapelle2015}.
What is particularly interesting about this companion is that \citet{Bowler2011} show that its spectral features strongly hint towards the presence of a circumplanetary disk (CPD).
They examined the possibility of a hypothetical scattering event by assuming it was formed much closer than its present location and was ejected to a wider orbit through a gravitational interaction with another massive body, and concluded that such an event would have likely disrupted the CPD.
Hence, they interpret the retention of the disk as evidence against a past scattering event, and argue that it was likely formed in-situ.
%
While in our case, \wisb may have followed a similar formation history and could have formed in-situ (e.g. by gravitational instability), it becomes increasingly difficult to explain an in-situ formation for an extremely wide separation planet such as \wisc through any currently known mechanism \citep[e.g.,][]{Veras2009,Nielsen2019}.

There aren't many known confirmed bound companions at projected orbital separations similar to that of \wisc, but a notable comparable object is Ross~458c, a 6-11~\mj companion orbiting at a projected separation of 1100~au from its host stars \citep{Goldman2010,Scholz2010,Burgasser2010}.
Although it may be more massive and could be orbiting at a larger semi-major axis than \wisc, a striking similarity between both systems is that Ross~458c also orbits a tight binary, Ross~458~AB.
This binary is a M0.5~Ve~+~M7~Ve pair \citep{Burgasser2010}, and, with a lower limit of $~\sim$150~Myr and an upper limit of $~\sim$800~Myr, is considerably older than \wisa.
Various analyses of the atmosphere of Ross~458c found that it is best described by a cloudy model, and showed that best-fit models incorporated sulphide clouds or silicate clouds \citep{Burningham2011,Morley2012,Manjavacas2019,Gaarn2023}.
\citet{Gaarn2023} derive a relatively high mass of $27\pm4$~\mj, well above the planetary-mass regime, leading to a correspondingly high mass ratio with its host binary, which they argue makes a planetary origin less likely.
However, they also show that its C/O ratio appears to be notably higher than that of its primaries and argue that this may suggest a planetary formation route for Ross~458c, followed by outward migration to its current position.
In the case of \wisc, its lower mass-ratio with the binary, combined with the presence of both a stellar binary companion and another planetary companion in the system, makes a similar formation scenario---initial formation closer in, followed by scattering to a wide orbit---a plausible origin.

Another system that is structurally analogous in several respects is YSES~1 \citep{Bohn2020,bohn2020b}, a young solar-type star that also hosts two directly imaged planetary companions.
Its estimated temperature of $4590\pm50$~K and age of $15\pm5$~Myr \citep{Bohn2020} are comparable to those of the primary of \wisa.
The colours and masses of the planets are also similar: \wisb is approximately 4~\mj less massive than YSES~1b and \wisc is roughly 2~\mj less massive than YSES~1c.
\citet{Zhang2024} found that the C/O ratio of YSES~1b is consistent with that of YSES~1, and argue that this may suggest (in-situ) formation via gravitational instability or core accretion beyond the snowline.
For YSES~1c, they derived the C/O ratio to be either solar, consistent with in-situ formation like YSES~1b, or sub-solar, consistent with formation within the CO iceline followed by outward scattering.
The YSES\,1 planets have been reported to show silicate features in their atmospheres, and a CPD has been directly detected around YSES\,1b \citep{Hoch2025}.
Like in the case of GSC~06214-00210b, if the CPD was formed during planet formation, it could further support the hypothesis of in-situ formation for YSES~1b.
\citet{Zhang2024} also argue that, given its low mass, YSES~1c is more likely to have formed closer to the star via bottom-up formation and was subsequently scattered outward to its current position.
This prompts the intriguing question of whether the two known planets in the system may have distinct formation pathways.
Given the similarities to the YSES~1 system, this raises the possibility that perhaps the \wisa planets, too, may have formed through different formation pathways.

\subsection{Wide-separation circumbinary planets}
While circumbinary planets are much more rare than planets orbiting single stars, there are a few notable examples that share similarities with the \wisa system.
A circumbinary planet similar in mass to \wisb is b~Cen~(AB)b, a $10.9\pm1.6$~\mj planet orbiting at $556\pm17$~au from b~Cen~AB \citep{Janson2021}.
Although with a mass of 6-10~\mj, the central pair is considerably more massive than that of \wisa, it is of a similar age and like \wisa, it was never part of any very high density environment.
The latter lowers the probability of gravitational capture scenarios, which was estimated to be approximately 17\% for b~Cen~(AB)b \citep{Janson2021}.
While scattering caused by the binary remains a possibility, the low eccentricity of b~Cen~(AB)b favours formation close to its present location, most likely due to gravitational instability processes.
Scattering is considered to be the more viable option for HD~106906~b \citep{Bailey2014}, an $~11$~\mj circumbinary planet located at a projected physical separation of approximately 737~au from its host.
Like \wisa, HD~106906 is a $\sim$15~Myr system that was initially thought to be a single star but was later revealed to be a close binary \citep{Lagrange2016}.
With a total stellar mass of at least $2.5$~\msun, this places its companion at a mass ratio close to that of \wisc.
Various formation mechanisms have been put forward for this planet, with suggestions of planet-capture, as well as in-situ formation through gravitational instability and formation closer in followed by outward scattering to a wider orbit.
While the relatively low eccentricity of 0.4 \citep{Nguyen2021} would seem to favour in-situ formation, simulations from \citet{Rodet2017} show that this configuration can be reproduced if the orbit of the companion scattered outward through interaction with the binary is stabilized by interaction with its own circumstellar disk, or by an external perturber through a fly-by.
Conversely, \citet{Moore2023} argue that the companion was most likely initially in a stable orbit and was instead scattered to its current configuration by a fly-by with a free-floating planet.
These examples illustrate that wide-orbit circumbinary planets may arise through various pathways, with both in-situ formation and formation closer to the stars followed by outward scattering remaining viable under certain conditions.
The similarities between these systems, in particular the wide-separation of the companions, suggest that the presence of a binary may play a role in the formation of such systems.

With a projected physical separation of at least 338~au, \wisb ranks among the top 10 widest-separation exoplanets with masses below 13~\mj; at 840~au, \wisc currently ranks seventh.\footnote{Retrieved from the NASA Exoplanet Archive on July 10, 2025: \url{https://exoplanetarchive.ipac.caltech.edu/}}
An interesting trend is that, of the ten currently known widest-separation ($\gtrsim$350~au) exoplanets, six \citep{Burgasser2010,Bailey2014,Dupuy2018,Janson2021,dupuy2023,Rothermich2024} orbit a stellar multiple system.
In contrast, among all exoplanets with a reported semi-major axis, fewer than 20\% orbit stellar multiples---a fraction that remains approximately the same when considering only directly imaged planets. 
Although the sample size is too small to allow for any definitive statistical conclusions, this trend raises the question of whether dynamical interactions in multiple systems may have played a role in forming companions at or scattering companions to such wide separations.
%
These objects might represent a distinct population of wide-separation giant planets; future large scale surveys and discoveries will be essential to better understand their formation pathways and dynamical evolution.

\section{Conclusions}
\label{sec:conclusion}

In this paper we present the detection and characterisation of two comoving planetary companions to a young stellar binary.
Based on lithium depletion isochrones, we derived the age of the primary to be $15.6_{-3.7}^{+4.1}$~Myr.
Using theoretical evolutionary models to convert the distance and photometry into luminosity and adopting this age estimate, the masses of the two companions are $10.4^{+1.1}_{-0.8}$ $\mathrm{M_J}$ and 
$5.3^{+1.1}_{-0.6}$ $\mathrm{M_J}$ at projected physical separations of 338 au and 840 au, respectively.
The astrometry across three epochs is inconsistent with a distant background source.
No orbital motion of the companions around the central binary is detected, within the errors on the measured astrometry.

Various formation pathways are possible for both planets, from in-situ formation via gravitational instability, to formation closer in followed by outward scattering, to planet-capture scenarios.
Constraining these possibilities requires detailed follow-up of their atmospheric chemistries and precise astrometric monitoring to determine their orbital periods and eccentricities.
Future work includes deeper photometry and spectroscopy of both planets to confirm their low mass and low gravity and potentially measure their rotation period.
Astrometric monitoring, especially of \wisb, with high precision instruments such as GRAVITY will help determine their orbital parameters.
Deeper ground based observations from large telescopes will enable searches for additional companions in the system, and JWST will be able to detect wider separation sub-Jupiter mass companions.
Together, these measurements will constrain the range of viable formation models for these exoplanets.

\begin{acknowledgements}

This publication makes use of VOSA, developed under the Spanish Virtual Observatory (\url{https://svo.cab.inta-csic.es}) project funded by MCIN/AEI/10.13039/501100011033/ through grant PID2020-112949GB-I00.
VOSA has been partially updated by using funding from the European Union's Horizon 2020 Research and Innovation Programme, under Grant Agreement nº 776403 (EXOPLANETS-A). 
This research has used the SIMBAD database, operated at CDS, Strasbourg, France \citep{wenger2000}.
This work has used data from the European Space Agency (ESA) mission {\it Gaia} (\url{https://www.cosmos.esa.int/gaia}), processed by the {\it Gaia} Data Processing and Analysis Consortium (DPAC, \url{https://www.cosmos.esa.int/web/gaia/dpac/consortium}).
Funding for the DPAC has been provided by national institutions, in particular the institutions participating in the {\it Gaia} Multilateral Agreement.
This research has made use of NASA's Astrophysics Data System.
Part of this research was carried out at the Jet Propulsion Laboratory, California Institute of Technology, under a contract with the National Aeronautics and Space Administration (80NM0018D0004).
This research made use of SAOImageDS9, a tool for data visualization supported by the Chandra X-ray Science Center (CXC) and the High Energy Astrophysics Science Archive Center (HEASARC) with support from the JWST Mission office at the Space Telescope Science Institute for 3D visualization \citep{2003ASPC..295..489J}.
The authors thank the anonymous reviewer for their swift and helpful feedback to this publication.
To achieve the scientific results presented in this article we made use of the \emph{Python} programming language\footnote{Python Software Foundation, \url{https://www.python.org/}}, including the \emph{SciPy} \citep{virtanen2020}, \emph{NumPy} \citep{numpy}, \emph{Matplotlib} \citep{Matplotlib}, \emph{emcee} \citep{ForemanMackey13}, \emph{astropy} \citep{Astropy2013,Astropy2018,Astropy2022}, \emph{backtracks} \citep{backtracks_zenodo} and \emph{PynPoint} \citep{Amara2012,Stolker2019}.
\end{acknowledgements}

%
\bibliographystyle{aa} 
\bibliography{main}

\begin{appendix}





\section{Median combined observations}
\label{app:obs}
\begin{figure}
\centering
\includegraphics[width=\hsize]{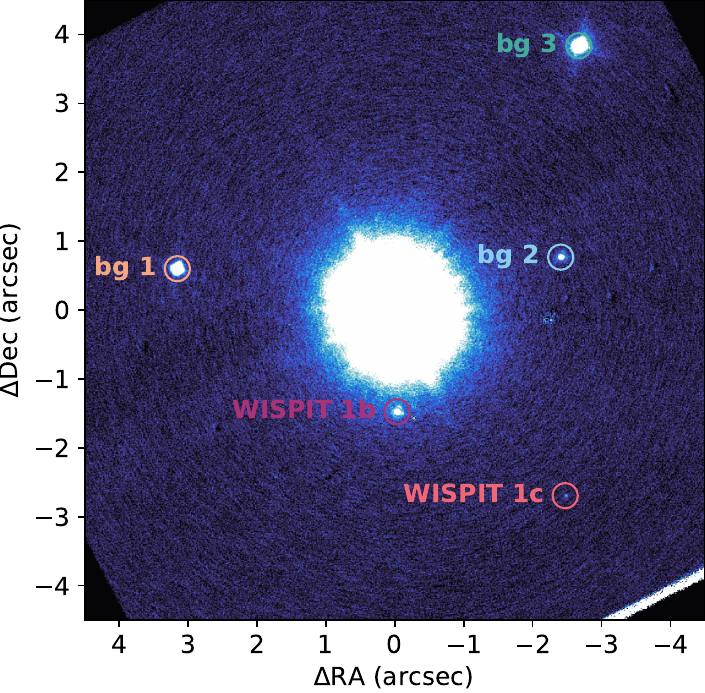}
\caption{\wisb, 1c, and background sources highlighted in the $H$-band median-combined image of epoch 2022-11-19.}
\label{fig:full_2022}
\end{figure}
\begin{figure}
\centering
\includegraphics[width=\hsize]{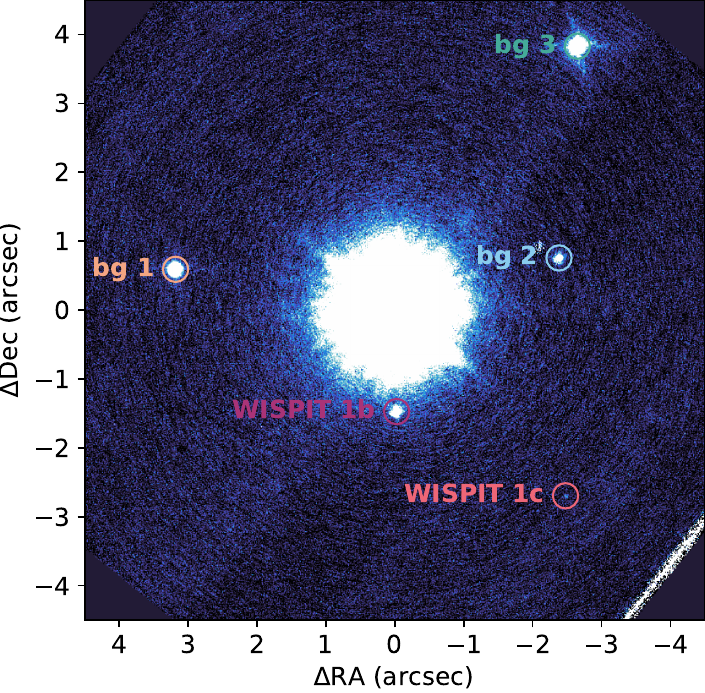}
\caption{\wisb, 1c, and background sources highlighted in the $H$-band median-combined image of epoch 2023-12-03.}
\label{fig:full_2023}
\end{figure}
\begin{figure}
\centering
\includegraphics[width=\hsize]{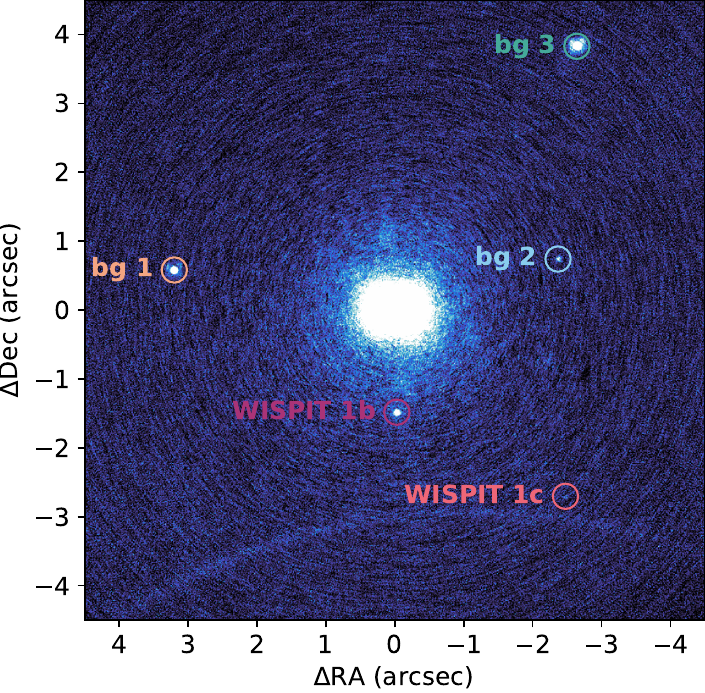}
\caption{\wisb, 1c, and background sources highlighted in the $K_s$-band median-combined image of epoch 2024-11-30.}
\label{fig:full_2024}
\end{figure}
The median combined observations from 2022, 2023, and 2024, with all detected sources in the field of view annotated, are shown in Figs. \ref{fig:full_2022}, \ref{fig:full_2023} and \ref{fig:full_2024}, respectively.
Of the $H$-band observations, the 2023 epoch (Fig.~\ref{fig:full_2023}) has better reported observing conditions and AO correction compared to 2022 (Fig.~\ref{fig:full_2022}), resulting in a higher SNR for sources closer to the star, such as \wisb.
However, the 2023 observation visually appears noisier overall, either due to increased detector noise or atmospheric effects. 
This is reflected in the background RMS, which is notably higher at larger radial separations in the 2023 data compared to 2022. 
Consequently, background-limited sources farther from the star, such as \wisc, have a slightly better SNR in the 2022 epoch.
The 2024 $K_s$-band observation (Fig.~\ref{fig:full_2024}) was taken under reasonably good observing conditions, but show a thermal background pattern affecting the lower half of the frame.
Since \wisc is near this region, its measured flux is strongly affected, and the local background estimate becomes less reliable, resulting in a low-SNR detection. 

\section{Environs of \wisa}
\label{app:environ}

The star has not been previously assigned membership to any stellar groups, but we discuss its environs.
\wisa is $\sim$4$^{\circ}$ SW of the Vel OB2 association (alias: Collinder 173) \citep[e.g.,][]{deZeeuw1999}, however its members are more distant and have smaller proper motions \citep[$\varpi$ = $2.85\pm0.08$ mas, $d$ $\simeq$ $351\pm10$\,pc, \mura, \mudec\, = $-6.20\pm0.44$, $9.08\pm0.64$ \masyr;][]{Beccari2018,Mendigutia2022}.
\citet{Beccari2018} finds Vel OB2 to have six subclusters with a range of ages of 10-30 Myr, including two well-studied clusters ($\gamma$ Vel, NGC 2547) and four new subgroups.
\wisa is 1$^{\circ}$.5 away from the Vel OB2
subgroup [BBJ2018] 2, however its distance and proper motion again differ significantly from the star: \citep[$\varpi$ = $2.42\pm0.05$ mas, $d$ $\simeq$ $413\pm8$\,pc, \mura, \mudec\, = $-5.41\pm0.19$, $8.18\pm0.15$ \masyr;][]{Beccari2018}, with the distances differing by $\sim$184 pc, and tangential velocities
differing by $\sim$17 \kms\, at the distance of \wisa.
We conclude that despite \wisa's projected proximity to Vel OB2 and its subgroups, its distance and kinematics are inconsistent with membership to that complex.

A neighboring star with a separation of 111'' appears to share the proper motion and parallax of \wisa.
2MASS J07512310-5008109 (UCAC4 200-015609, Gaia DR3 5517037503494641024, GALAH 151231003201128) is a Li-rich star in the GALAH survey \citep[][] {Buder2018,Buder2021}, with detectable X-ray emission in the ROSAT and eROSITA X-ray All-Sky Surveys \citep[1RXS J075123.3-500756, 2RXS J075123.1-500754, 1eRASS J075122.9-500810;][]{Voges2000,Boller2016,Merloni2024,Freund2024}.
Naturally, the star appears to be fast-rotating, with the GALAH survey
reporting $v \sin{i}$ = $20.812\pm1.018$ \kms{} \citep{Buder2018}, and \citet{Green2023} estimated $P_{\rm rot}$ = 5.47\,d using TESS time series photometry.
Using Gaia DR3 astrometry and epoch 2016.0 positions, we estimate 2MASS J07512310-5008109 to be at separation $\rho$ = $109.886066\pm0.000022$ arcsec and PA = 87$^{\circ}$.420226 with respect to \wisa, translating
to projected separation $\Delta$ = 25147\,au at $d$ = 229\,pc. 
Their proper motions differ by $\Delta$\mura, $\Delta$\mudec\, = $0.283\pm0.020$, $2.465\pm0.021$ \masyr, which at $d$ = 229\,pc translates to difference in tangential velocities of $\Delta V_{\alpha}$, $\Delta V_{\delta}$ = $0.307\pm0.022$, $2.674\pm0.023$ \kms.
The Gaia DR3 parallaxes for \wisa and 2MASS J07512310-5008109 are in remarkable agreement -- within 1.45$\sigma$ ($\Delta$$\varpi$ = $0.0246\pm0.0170$ mas), and their inferred Gaia DR3 distances
differ by only $\Delta$$d$ = $0.356\pm0.810$ pc \citep{BailerJones2021}. 
2MASS J07512310-5008109 is brighter ($G$ = 11.72, $V$ = 11.91) than \wisa ($G$ = 12.72, $V$ = 13.15), and hotter \citep[$T_{\rm eff}$ $\simeq$ 5730\,K;][]{GaiaDR3} so we suspect that 2MASS J07512310-5008109 may be the primary of this system.
Another very faint and very red comoving codistant star in \citet{Zari2018} is 2MASS J07485619-4656229. This star has a parallax of $\varpi$ = $4.43 \pm 0.08$ mas ($d$ $\simeq$ $227\pm8$\,pc) and a proper motion of \mura, \mudec\, = $-20.766\pm0.096$, $16.099\pm0.098$ \masyr.
The age assigned to this star by \citet{Kerr2021} is $16.8^{+3.5}_{-3.9}$ Myr.


\section{Reddening towards \wisa}
\label{app:reddening_extinction}
\begin{figure}
\centering
\includegraphics[width=\hsize]{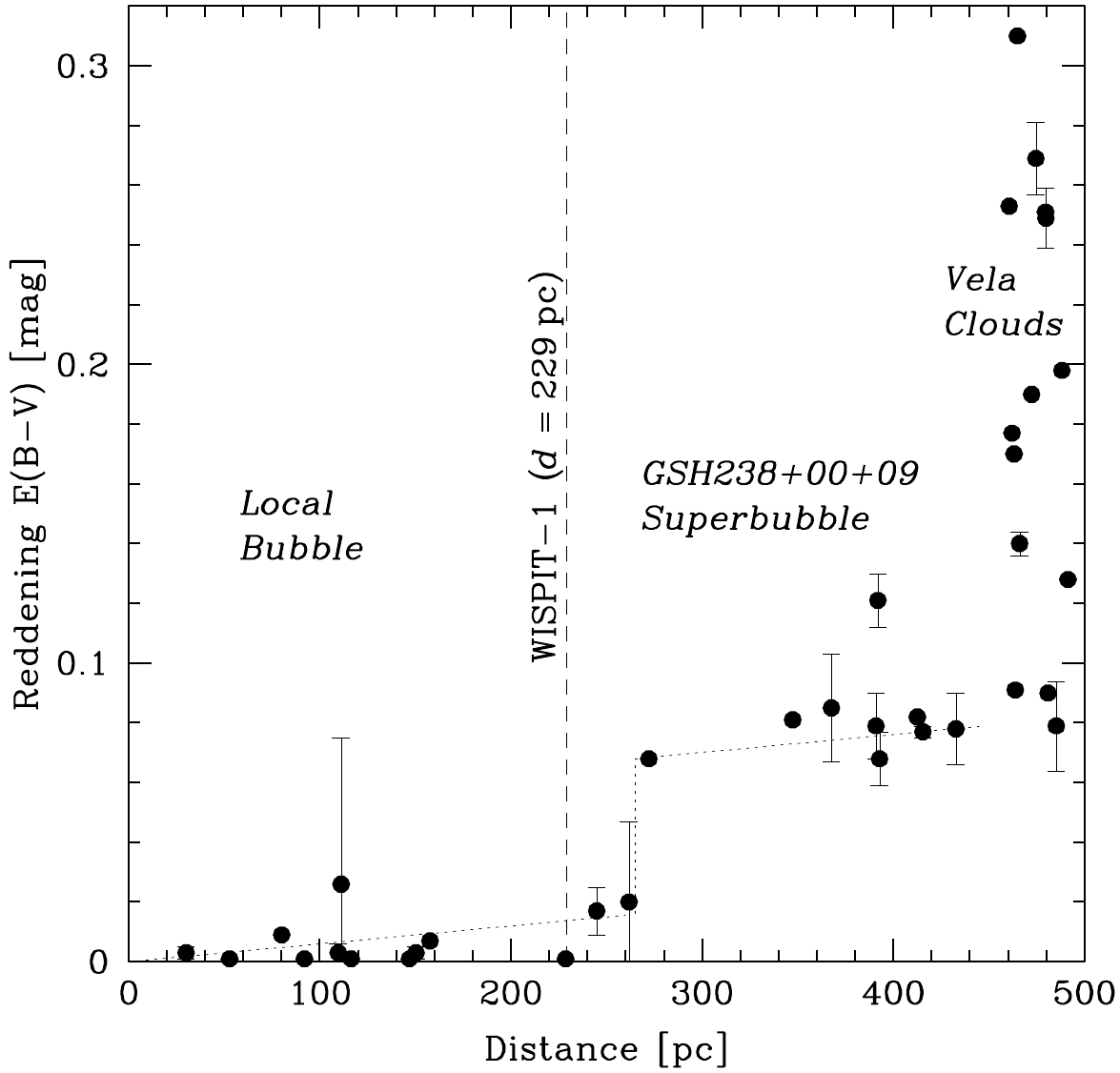}
\caption{Distance versus reddening \EBV\, for stars within 2$^{\circ}$ of \wisa.
Distances are calculated as $d$ = 1/$\varpi$ from the default parallaxes in SIMBAD, and the reddening \EBV\, values are the mean values provided by \citet{Paunzen2024}.
The long-dashed vertical list corresponds to the distance to \wisa ($d$ = 229\,pc), and the dotted lines are fiducial \EBV\, reddening slopes of 0.06 mag\,kpc$^{-1}$ up to distance $d$ $\simeq$ 265\,pc, where there is a \EBV\,$\simeq$\,0.05 mag wall of reddening at the interface between the Local Bubble and GSH238+00+09 Super-Bubble \citep{Heiles1998,Lallement2014,O'Neill2024}.}
\label{fig:ebv}
\end{figure}

Gaia DR3 quotes unusually high reddening and extinction estimates
for \wisa: 
$A_0$ = $1.4306^{+0.0030}_{-0.0033}$ mag,
$A_G$ = $1.1341^{+0.0025}_{-0.0027}$ mag,
$E(B_p - R_p)$ = $0.6128^{+0.0014}_{-0.0016}$.
These are unexpected as \wisa does not appear to be projected towards a particularly dusty region, and the maximum reddening expected from interstellar dust is predicted to be $E(B-V)_{max}$ = $0.2160\pm0.0162$ ($A_{V max}$ = $0.6482$ mag) based on the dust maps from \citet{Schlafly2011}\footnote{Queried via the Galactic Dust Reddening and Extinction tool at IRSA \url{https://irsa.ipac.caltech.edu/applications/DUST/}, with the mean and rms based on pixels within 5' radius.}. 
Given the importance of extinction to the assessment of the intrinsic colours, effective temperatures, luminosities, and inferred ages for \wisa and its companions, we investigate the reddening in more detail.

To get an idea of the expected reddening due to interstellar medium towards \wisa, we queried the \EBV{} reddening values for stars from \citet{Paunzen2024}, and cross-reference their 2MASS IDs with SIMBAD to assign parallaxes (usually from Gaia DR3) and inferred distances. 
In Fig.~\ref{fig:ebv}, we plot the distances of stars within 2$^{\circ}$ of \wisa (calculated as $d$ = 1/$\varpi$) with \EBV\, values in the \citet{Paunzen2024} catalog. 
The general trend of reddening in the direction of \wisa can be characterized as follows: (i) negligible reddening within $d$ $\lesssim$ 265\,pc, with -- at most -- a reddening trend of $\Delta E(B-V)$/$\Delta d$ $\simeq$ 0.06\,mag\,kpc$^{-1}$.
(ii) A ``wall'' of reddening $\Delta E(B-V)$ $\simeq$ 0.05 mag at $d$ $\simeq$ 265\,pc, beyond which all of the neighbouring stars have at least $E(B-V)$ $\geq$ 0.07\,mag.
(iii) Another low-density region at distances $\sim$265\,pc $\lesssim$ $d$ $\lesssim$ 50 with similar reddening slope, followed by a region of larger \EBV\, reddening values ($\sim$0.1-0.3 mag) with large scatter beyond $d$ $>$ 450\,pc.

A larger scale view of the 3D reddening can be viewed in Fig.~1 of \citet{Lallement2014}, where \wisa ($\ell, b$ = 263$^{\circ}$.70, -11$^{\circ}$.71, $d$ = 229\,pc) sits at (X, Y $\simeq$ -25, -224\,pc).
On the scales of tens of degrees, the Local Bubble was previously known to stretch to $d$ $\simeq$ 200-300\,pc in the general direction of \wisa \citep[e.g., Fig. 3 of ][]{O'Neill2024}. 
The ``wall'' of reddening \EBV\, $\simeq$ 0.05 mag at $d$ $\simeq$ 265\,pc appears to be the interface between the Local Bubble and the GSH238+00+09 super-bubble \citep{Heiles1998,Lallement2014}.
The \citet{Paunzen2024} catalog contains some neighboring hot stars at similar distances as \wisa with very small reddenings, e.g., the B9V star HD 66192 (2MASS J08000557-4854195, 114' away, $d$ = 228\,pc) with $E(B-V)$ = 0.001 mag, the B2V star HD 64740 (2MASS J07530364-4936469, 36' away, $d$ = 245\,pc) with $E(B-V)$ = $0.017\pm0.008$ mag, the F2III/IV star HD 63176 (2MASS J07450716-5017163, 59' away, $d$ = 262\,pc) with $E(B-V)$ = $0.027\pm0.027$ mag. 
The TIC \citep{Stassun2019} predicts $E(B-V)$ = $0.0674\pm0.0103$ mag based on a very simple model assuming an exponential model of disk reddening and maximum reddening informed by the \citet{Schlegel1998} dust maps, but the estimated value seems too high (more appropriate for $d$ $>$ 275\,pc stars) and does not account for the obvious bubbles and their interfaces observed in the reddening estimates (e.g., Fig.~1). 
So we discount the TIC reddening estimate. 
The trend seen among the neighboring stars with \citet{Paunzen2024} reddening estimates suggests that WISPIT is just within the inner edge of the Local Bubble with interstellar reddening of approximately $E(B-V)$ $\simeq$ $0.015\pm0.015$ mag ($A_V$ $\simeq$ $0.047\pm0.047$ mag).
We conclude that \wisa lies within the Local Bubble, and there is no evidence for large scale {\it interstellar} dust in the 3D vicinity or foreground of \wisa that can account for the large reddenings and extinctions quoted by \citet{GaiaDR3}.



\section{Apparent magnitude decomposition}
\label{app:mag_decomposition}

To derive the apparent magnitudes of the primary and secondary stars from the observed 2MASS system magnitudes, we convert 2MASS total magnitudes to fluxes, apply the measured flux ratios, and convert back to magnitudes. 
In the derivation included below, $H_{\mathrm{tot}}$ and $K_{s,\mathrm{tot}}$ denote the total magnitudes from Table~\ref{tab:star}, and $R_H$, $R_{K_s}$ the secondary-to-primary flux ratios in each band. 
The primary is denoted with $A$ and the secondary with $B$. 
The associated uncertainties are computed with standard error propagation of the uncertainties in the flux ratio and the 2MASS magnitudes.

\begin{alignat*}{3}
&\textbf{Primary:} \quad
&F_H^{A} &= \frac{10^{-0.4 H_{\mathrm{tot}}}}{1 + R_H}, \quad
&F_{K_s}^{A} &= \frac{10^{-0.4 K_{s,\mathrm{tot}}}}{1 + R_{K_s}}, \\
&&H_A &= -2.5 \log_{10}(F_H^{A}), \quad
&K_{s,A} &= -2.5 \log_{10}(F_{K_s}^{A}). \\
\\[-1.2ex]
&\textbf{Secondary:} \quad
&F_H^{B} &= 10^{-0.4 H_{\mathrm{tot}}} - F_H^{A}, \quad
&F_{K_s}^{B} &= 10^{-0.4 K_{s,\mathrm{tot}}} - F_{K_s}^{A}, \\
&&H_B &= -2.5 \log_{10}(F_H^{B}), \quad
&K_{s,B} &= -2.5 \log_{10}(F_{K_s}^{B}).
\end{alignat*}

\section{Lithium isochrones}
\label{app:isochrones}
\begin{figure}[h]
   \centering
   \includegraphics[width=\hsize]{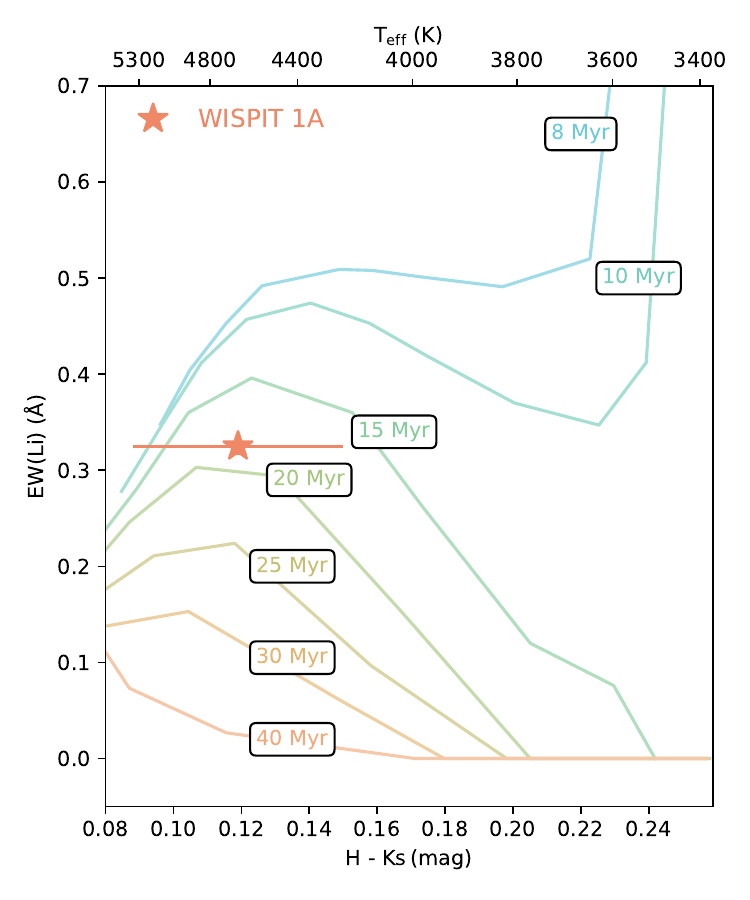}
      \caption{Lithium isochrones from \citet{Zerjal2021} adapted to $H-K_s$. WISPIT 1A is indicated with an orange star.}
         \label{fig:isochrones}
\end{figure}

Fig. \ref{fig:isochrones} shows lithium equivalent width as a function of $H-K_s$ colour for various stellar ages. Note that the flux at the \ion{Li}{I} line (6708 $\AA$) is dominated by the primary star; the much fainter and cooler secondary star is not expected to contribute to the lithium absorption measurement.

\section{Proper motion of background objects}
\label{app:background_ccs}
\begin{figure*}[h!]
    \centering
    \begin{subfigure}[t]{0.5\textwidth}
        \centering
        \includegraphics[width=0.99\textwidth]{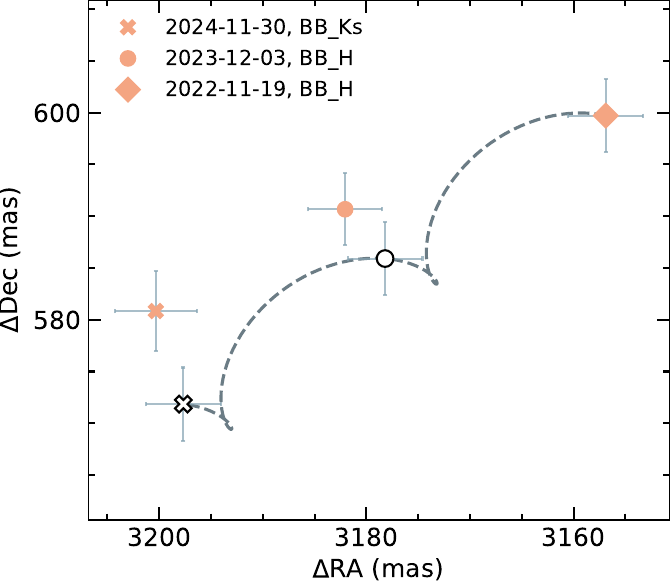}
        \caption{Background object 1}
    \end{subfigure}%
    ~ 
    \begin{subfigure}[t]{0.5\textwidth}
        \centering
        \includegraphics[width=0.99\textwidth]{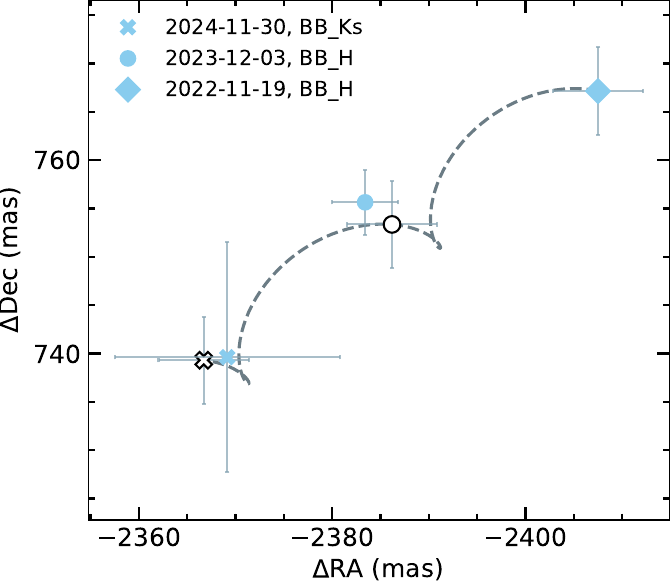}
        \caption{Background object 2}
    \end{subfigure}
    ~
    \begin{subfigure}[t]{0.5\textwidth}
        \centering
        \includegraphics[width=0.99\textwidth]{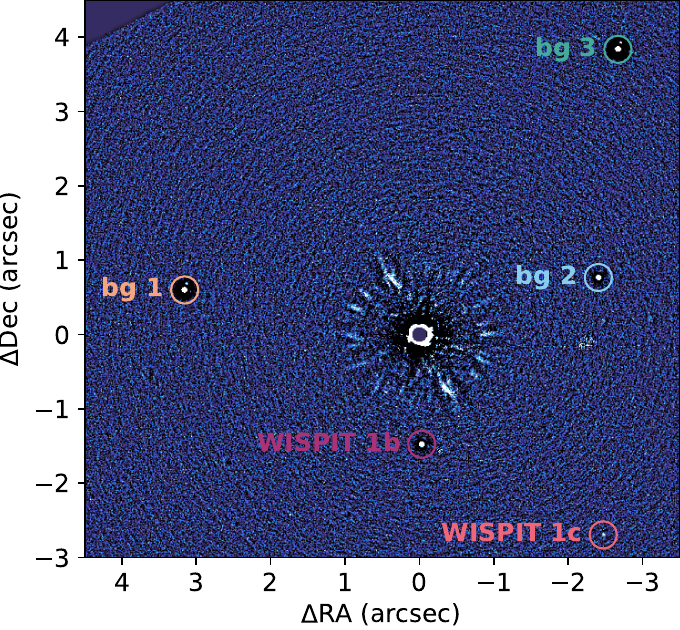}
        \caption{Detection map}
    \end{subfigure}%
    ~ 
    \begin{subfigure}[t]{0.5\textwidth}
        \centering
        \includegraphics[width=0.99\textwidth]{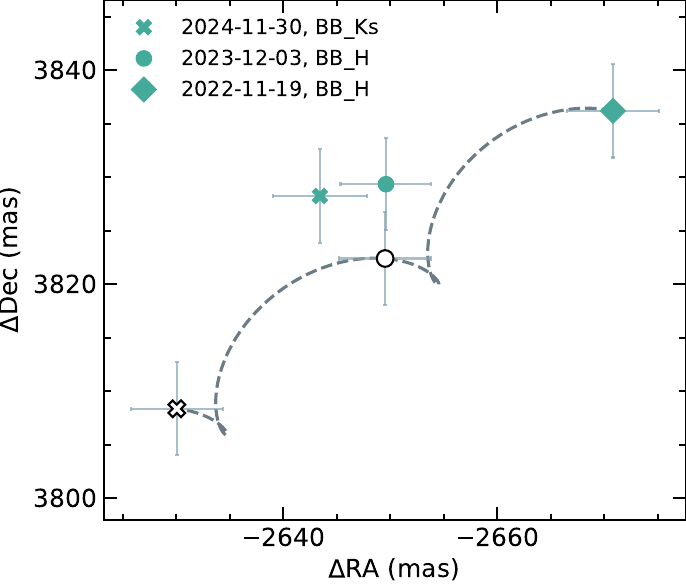}
        \caption{Background object 3}
    \end{subfigure}
    
    \caption{Proper motion analysis of all background sources in the field of view. The detection map in panel (c) shows the positions of the sources relative to \wisa in the observation of epoch 2022-10-19 processed with unsharp masking. The three background sources are shown in panels (a), (b), and (d), respectively. Each epoch is marked with a unique marker shape; diamond (2022), circle (2023) and cross (2024). The coloured version of the marker indicates the measured position of the companion. The unfilled (black outline, white center) version of this marker indicates the position that the companion would have been at at the corresponding date if it were a background object. The dashed curve depicts the parallactic track of a stationary background object from first epoch to last epoch. The $\chi^2$ between the measured positions of background objects 1, 2 and 3 are 10.6, 1.1 and 32.2, respectively.}
    \label{fig:background_ppm}
\end{figure*}

Fig.~\ref{fig:background_ppm} presents the measured positions of the background sources within the field of view, along with predicted background tracks for stationary background objects. It is evident from this figure that the astrometry of these sources is consistent with stationary background objects, which strengthens the interpretation of \wisb and \wisc as comoving planetary companions, since they clearly deviate from these predicted tracks. We note, however, that the third epoch position of background object~3 does not follow the expected trajectory. Although it is possible that it is a non-stationary background source, another possible explanation is that this source may be a binary, as its PSF visually slightly deviates from that of a single point-source.
\onecolumn



\FloatBarrier
\section{Median PSFs}
\label{app:median_psfs}
All observations used to create the normalized median flux PSF in $H$-band and $K_s$-band are listed in Tables~\ref{tab:med_psf_H} and ~\ref{tab:med_psf_Ks}, respectively.
Here `Target name' denotes the commonly used designation for the source, and `Archive name' corresponds to the name it is registered under in the ESO Science Archive Facility.

\def\arraystretch{1.2}
\setlength{\tabcolsep}{55pt}
\begin{longtable}{@{}llr@{}}
\caption{Flux observations used for constructing the median $H$-band PSF.} \label{tab:med_psf_H} \\
\hline\hline
Target name & Archive name & Observation date \\
\hline
\endfirsthead

\hline
Target name & Archive name & Observation date \\
\hline
\endhead

\hline
\endfoot

\hline
\endlastfoot
1RXS J114519.6-574925 & 2MASS J11452016-5749094 & 2018-05-14 \\
1RXS J114542.7-573928 & 2MASS J11454278-5739285 & 2019-01-13 \\
1RXS J114542.7-573928 & 2MASS J11454278-5739285 & 2023-04-20 \\
1RXS J121010.3-485538 & 2MASS J12101065-4855476 & 2017-04-18 \\
1RXS J123834.9-591645 & 2MASS J12383556-5916438 & 2019-01-03 \\
1RXS J123834.9-591645 & 2MASS J12383556-5916438 & 2019-01-12 \\
1RXS J123834.9-591645 & 2MASS J12383556-5916438 & 2023-07-14 \\
1RXS J124830.1-594449 & 2MASS J12483152-5944493 & 2023-08-07 \\
2MASS J12374883-5209463 & 2MASS J12374883-5209463 & 2018-12-30 \\
2MASS J12374883-5209463 & 2MASS J12374883-5209463 & 2023-07-14 \\
2MASS J13065439-4541313 & 2MASS J13065439-4541313 & 2018-04-08 \\
2MASS J13065439-4541313 & 2MASS J13065439-4541313 & 2023-07-08 \\
2MASS J13121764-5508258 & 2MASS J13121764-5508258 & 2018-05-15 \\
2MASS J13334410-6359345 & 2MASS J13334410-6359345 & 2023-06-16 \\
ASAS J114452-6438.9 & 2MASS J11445217-6438548 & 2018-05-14 \\
ASAS J114452-6438.9 & 2MASS J11445217-6438548 & 2023-04-20 \\
ASAS J122648-5214.8 & 2MASS J12264842-5215070 & 2018-12-30 \\
ASAS J122648-5214.8 & 2MASS J12264842-5215070 & 2023-05-28 \\
ASAS J124547-5411.0 & 2MASS J12454884-5410583 & 2018-04-30 \\
ASAS J130550-5304.2 & 2MASS J13055087-5304181 & 2022-04-02 \\
ASAS J131033-4816.9 & 2MASS J13103245-4817036 & 2018-05-01 \\
CD-41 7947 & 2MASS J13343188-4209305 & 2023-08-07 \\
CD-47 7559 & 2MASS J12220430-4841248 & 2017-04-18 \\
CD-51 6900 & 2MASS J12404664-5211046 & 2018-04-30 \\
CD-51 6900 & 2MASS J12404664-5211046 & 2023-05-30 \\
CD-51 7268 & 2MASS J13064012-5159386 & 2018-04-30 \\
CD-51 7268 & 2MASS J13064012-5159386 & 2023-06-15 \\
CD-57 4328 & 2MASS J12113142-5816533 & 2018-12-22 \\
CD-57 4328 & 2MASS J12113142-5816533 & 2019-02-18 \\
CPD-50 5313 & 2MASS J12361767-5042421 & 2018-12-30 \\
CPD-52 6110 & 2MASS J13015069-5304581 & 2019-01-08 \\
CPD-56 5307 & 2MASS J12333381-5714066 & 2019-01-01 \\
CPD-56 5307 & 2MASS J12333381-5714066 & 2019-01-14 \\
CPD-56 5307 & 2MASS J12333381-5714066 & 2023-05-28 \\
CPD-64 1859 & 2MASS J12192161-6454101 & 2023-06-17 \\
PM J12160-5614 & 2MASS J12160114-5614068 & 2018-12-27 \\
RX J1216.6-7007A & 2MASS J12164023-7007361 & 2018-12-23 \\
RX J1216.6-7007A & 2MASS J12164023-7007361 & 2019-02-15 \\
RX J1216.6-7007A & 2MASS J12164023-7007361 & 2023-12-21 \\
RX J1216.6-7007A & TYC 9231-1185-1 & 2024-06-10 \\
RX J1220.0-5018A & 2MASS J12195938-5018404 & 2018-12-30 \\
RX J1220.0-5018A & 2MASS J12195938-5018404 & 2023-06-17 \\
RX J1230.5-5222 & 2MASS J12302957-5222269 & 2018-12-30 \\
RX J1230.5-5222 & 2MASS J12302957-5222269 & 2022-03-30 \\
UCAC2 12444765 & 2MASS J13095880-4527388 & 2018-05-01 \\
UCAC4 186-087394 & UCAC4 186-087394 & 2024-06-13 \\
UNSW-V 514 & 2MASS J13174687-4456534 & 2018-05-28 \\
V1257 Cen & 2MASS J12505143-5156353 & 2019-01-12 \\
\end{longtable}

\def\arraystretch{1.2}
\setlength{\tabcolsep}{55pt}
\begin{longtable}{@{}llr@{}}
\caption{Flux observations used for constructing the median $K_s$-band PSF.}
\label{tab:med_psf_Ks} \\
\hline\hline
Target name & Archive name & Observation date \\
\hline
\endfirsthead

\hline
Target name & Archive name & Observation date \\
\hline
\endhead

\hline
\endfoot

\hline
\endlastfoot
1RXS J114542.7-573928 & 2MASS J11454278-5739285 & 2019-01-13 \\
1RXS J123834.9-591645 & 2MASS J12383556-5916438 & 2019-01-03 \\
1RXS J123834.9-591645 & 2MASS J12383556-5916438 & 2019-01-12 \\
2MASS J12374883-5209463 & 2MASS J12374883-5209463 & 2018-12-30 \\
2MASS J13065439-4541313 & 2MASS J13065439-4541313 & 2018-04-08 \\
2MASS J13121764-5508258 & 2MASS J13121764-5508258 & 2018-05-15 \\
ASAS J114452-6438.9 & 2MASS J11445217-6438548 & 2018-05-14 \\
ASAS J122105-7116.9 & 2MASS J12210499-7116493 & 2019-01-12 \\
ASAS J122648-5214.8 & 2MASS J12264842-5215070 & 2018-12-30 \\
ASAS J124547-5411.0 & 2MASS J12454884-5410583 & 2018-04-30 \\
ASAS J130550-5304.2 & 2MASS J13055087-5304181 & 2018-06-11 \\
ASAS J130550-5304.2 & 2MASS J13055087-5304181 & 2018-07-05 \\
ASAS J131033-4816.9 & 2MASS J13103245-4817036 & 2018-05-01 \\
CD-49 7280 & 2MASS J12405458-5031550 & 2018-12-30 \\
CD-51 6900 & 2MASS J12404664-5211046 & 2018-04-30 \\
CD-57 4328 & 2MASS J12113142-5816533 & 2018-12-22 \\
CD-57 4328 & 2MASS J12113142-5816533 & 2019-02-18 \\
CPD-49 4947 & 2MASS J12121119-4950081 & 2018-12-22 \\
CPD-50 5313 & 2MASS J12361767-5042421 & 2018-12-30 \\
CPD-53 5235 & 2MASS J12365895-5412178 & 2019-01-01 \\
CPD-53 5235 & 2MASS J12365895-5412178 & 2019-01-13 \\
CPD-56 5307 & 2MASS J12333381-5714066 & 2019-01-01 \\
CPD-56 5307 & 2MASS J12333381-5714066 & 2019-01-14 \\
PM J12160-5614 & 2MASS J12160114-5614068 & 2018-12-27 \\
RX J1216.6-7007A & 2MASS J12164023-7007361 & 2018-12-23 \\
RX J1216.6-7007A & 2MASS J12164023-7007361 & 2019-02-15 \\
RX J1220.0-5018A & 2MASS J12195938-5018404 & 2018-12-30 \\
RX J1230.5-5222 & 2MASS J12302957-5222269 & 2018-12-30 \\
UCAC2 12444765 & 2MASS J13095880-4527388 & 2018-05-01 \\
UCAC4 186-087394 & 2MASS J12510556-5253121 & 2019-01-08 \\
UNSW-V 514 & 2MASS J13174687-4456534 & 2018-05-28 \\
V1257 Cen & 2MASS J12505143-5156353 & 2019-01-12 \\
\end{longtable}

\FloatBarrier 
\clearpage

\end{appendix}
\end{document}